\documentclass[14pt]{article}
\pdfoutput=1
\usepackage{amsmath,amssymb,amsfonts,amsthm,bm,bbm,cancel,wasysym}
\usepackage{epsfig,graphics,graphicx,epstopdf}
\usepackage{array,booktabs,colortbl,colordvi,multirow}
\usepackage{colordvi,color,xcolor}
\usepackage{hyperref}
\usepackage{rotating}

\usepackage{verbatim}
\usepackage{cite}
\usepackage{subfig}
\usepackage{setspace}
\usepackage{url}
\usepackage[percent]{overpic}
\usepackage{slashed}
\usepackage{authblk}
\usepackage{xspace}
\usepackage{fullpage}
\usepackage{hyperref}

\def\ie{{\it i.e.}}
\def\eg{{\it e.g.}}
\def\etc{{\it etc}}

\newcommand{\qslash}{\cancel{q}~}

\def\to{\rightarrow}

\newskip\zatskip \zatskip=0pt plus0pt minus0pt
\def\matth{\mathsurround=0pt}
\def\lsim{\mathrel{\mathpalette\atversim<}}
\def\gsim{\mathrel{\mathpalette\atversim>}}
\def\atversim#1#2{\lower0.7ex\vbox{\baselineskip\zatskip\lineskip\zatskip
  \lineskiplimit 0pt\ialign{$\matth#1\hfil##\hfil$\crcr#2\crcr\sim\crcr}}}

\begin{document}

%----------------------------------- TITLE AND AUTHORS -----------------------------------------%

%Preprint numbers
\begin{flushright}
SLAC-PUB-17751\\
\today
\end{flushright}
\vspace*{5mm}

\renewcommand{\thefootnote}{\fnsymbol{footnote}}
\setcounter{footnote}{1}

\begin{center}

{\Large {\bf Towards UV-Models of Kinetic Mixing and Portal Matter V: Indirect Probes of the New Physics Scale
%in $[SU(3)]^4$ 
}}\\
%\vspace*{0.15cm}

\vspace*{0.75cm}

{\bf Thomas G. Rizzo}~\footnote{rizzo@slac.stanford.edu}

\vspace{0.5cm}

{SLAC National Accelerator Laboratory}\\ 
{2575 Sand Hill Rd., Menlo Park, CA, 94025 USA}

\end{center}
\vspace{.5cm}

%--------------------------------------------- ABSTRACT ---------------------------------------------%

\begin{abstract}
\noindent  

Kinetic mixing of the dark photon, the gauge boson of a hidden $U(1)_D$, with the Standard Model (SM) gauge fields to induce an interaction between ordinary matter and dark matter (DM) at 1-loop 
requires the existence of portal matter (PM) fields having both dark and SM charges. As discussed in earlier work, these same PM fields can also lead to other loop-level mechanisms besides kinetic 
mixing that can generate significant interactions between SM fermions and the dark photon in a manner analogous to those that can be generated between a Dirac neutrino and a SM photon, \ie, 
dark moments. In either case, there are reasons to believe, \eg, due to the RGE running of the $U(1)_D$ gauge coupling, that PM fields may have $\sim$ TeV-scale masses that lie at or above those 
directly accessible to the HL-LHC. If they lie above the reach of the HL-LHC, then the only way to possibly explore the physics at this high scale in the short term is via indirect measurements made at 
lower energies, \eg, at lepton colliders operating in the $m_Z$ to 1 TeV range.  In particular, processes such as $e^+e^- \to \gamma+$DM or $e^+e^-\to \bar ff$, where $f$ is a SM fermion, may be 
most useful in this regard. Here we explore these possibilities within the framework of a simple toy PM model, introduced in earlier work, based on a non-abelian dark gauge group completion operating 
at the PM scale. In the KM setup, we show these efforts fail due to the inherently tiny cross sections in the face of substantial SM backgrounds.  However, in the case of interactions via induced dark 
moments, since they necessarily take the form of higher dimensional operators whose influence grows with energy, we show that access to PM-scale information may become possible for certain ranges 
of the toy model parameters for both of these $e^+e^-$ processes at a 1 TeV collider.

\end{abstract}

\vspace{0.5cm}
\renewcommand{\thefootnote}{\arabic{footnote}}
\setcounter{footnote}{0}
\thispagestyle{empty}
\vfill
\newpage
\setcounter{page}{1}

%-------------------------------- DOCUMENT: INTRODUCTION ---------------------------------%

% 1 Introduction

\section{Introduction and Overview}

The Standard Model (SM) has been very successful at describing experimental observations made over a very wide range of both energy and physical scales. Perhaps it's greatest challenge is that 
it does not explain the nature of dark matter (DM), for which no SM candidates exist, and also how such DM may interact us, if at all, in a non-gravitational manner. This additional new interaction, which 
is expected to be quite weak (if actually present), is more than likely to be necessary for the DM to achieve the value of its relic mass density as measured by the Planck 
Collaboration\cite{Planck:2018vyg}. Possible models of what particle DM may be now go back some decades, the traditional candidates being the 
QCD axion\cite{Kawasaki:2013ae,Graham:2015ouw,Irastorza:2018dyq} or a possible member 
of the family of weakly interacting massive particles, \ie, WIMPS\cite{Arcadi:2017kky,Roszkowski:2017nbc}. Searches for both of these types of DM candidates continue to dive ever more deeply 
into regions of smaller cross section and over an ever widening range of particle masses as experimental sensitivities increase.  While either one (or both !) of these types of particles may yet be 
found, the many searches over the years have so far produced only negative results, be they direct or indirect detection experiments or via the on-shell DM production at 
the LHC\cite{LHC,Aprile:2018dbl,Fermi-LAT:2016uux,Amole:2019fdf,LZ:2022ufs}, with the result being that a respectable fraction of many of the proposed models' parameter spaces have now 
been excluded. 

In parallel with these ongoing searches, it has become quite clear that the set of possible particle DM models is vastly greater than what had been previously anticipated. Hypothetical DM candidates 
now cover extremely broad ranges in both masses as well as coupling strengths to at least some of the fields of 
the SM\cite{Alexander:2016aln,Battaglieri:2017aum,Bertone:2018krk,Cooley:2022ufh,Boveia:2022syt,Schuster:2021mlr}. To make even partially adequate coverage of this now huge parameter 
space will require wide varieties of different types of experiments which will aim at various subregions of mass and coupling strength lying within it. 
In order to broadly describe the nature of the possible interactions between DM and the SM it is possible to construct effective field theories (EFT), called `portals',  involving interactions which 
can be further classified as being either renormalizable (\ie, having mass dimension $\leq 4$) or non-renormalizable (\ie, with dimension $>4$){\footnote {Of course, such 
non-renormaliazble interactions may themselves simply be the result of integrating out very massive fields in a more UV-complete picture. This may also be true as well in the case of some of 
the renormalizable portals.}}  In either case, in addition to the DM and SM fields that we wish to connect, the various portals scenarios frequently require the existence of a variety of 
new mediator fields which are specifically designed to link both the visible and dark sectors in rather well-defined manners.

Of special interest to us here is the renormalizable kinetic mixing (KM)/vector portal\cite{KM,vectorportal,Gherghetta:2019coi} which has gotten much attention in the recent literature. One reason 
for this is that for an interesting range of model parameters, the DM can be a thermal relic\cite{Steigman:2015hda,Saikawa:2020swg} while still avoiding numerous other experimental constraints. 
This scenario envisions a new gauge boson, the dark photon (DP)\cite{Fabbrichesi:2020wbt,Graham:2021ggy}, $V$, associated with a hidden abelian gauge group, $U(1)_D$, (generally assumed to 
be broken, as in the SM, by the vev(s) of one or more dark 
Higgs fields) having a corresponding gauge coupling, $g_D$. It this type of setup, unlike in the familiar WIMP models, the thermal freeze-out of the DM yielding 
the observed relic density will now occur only for comparable, sub-GeV masses for the DM and DP and then only if the strength of the DM-SM interaction is of the correct magnitude. 
In this class of models, the SM fields are assumed to be neutral under this new $U(1)_D$ gauge group, \ie, they carry zero dark charge, $Q_D$, while the DM carries a non-zero charge. The 
desired SM-DM interaction, which needs to be quite weak, one imagines happening at the loop-level through the existence of some {\it additional} new heavy particles, here called Portal Matter (PM)\cite{Rizzo:2018vlb,Rueter:2019wdf,Kim:2019oyh,Rueter:2020qhf,Wojcik:2020wgm,Rizzo:2021lob,Rizzo:2022qan,Wojcik:2022rtk,Rizzo:2022jti,Rizzo:2022lpm,Wojcik:2022woa,Carvunis:2022yur,Verma:2022nyd,Rizzo:2023qbj,Wojcik:2023ggt,Rizzo:2023kvy}, carrying non-zero values of both the $U(1)_D$ as well as SM charges. The KM of the DP with the SM photon can then be described 
in the IR limit as the result of vacuum polarization-like graphs in which the PM fields participate; the strength of this kinetic mixing at such low scales is then given by a dimensionless parameter 
which is obtained by considering the sum
\begin{equation}
\epsilon =\frac{g_D  e}{24\pi^2} \sum_i ~\big(\eta_i   N_{c_i}Q_{em,i}Q_{D_i}\big)~ ln \frac{m^2_i}{\mu^2}\,,
\end{equation}
where $e$ is the usual QED gauge coupling and $m_i(Q_{em,i},Q_{D_i}, N_{c_i})$ are the mass (electric charge, dark charge, number of colors) of the $i^{th}$ PM field running in the loop.
Also here $\eta_i=1(1/2)$ if the PM particle running in the loop is assumed to be a chiral fermion (complex scalar). From this expression we see that if the condition 
\begin{equation}
\sum_i ~\eta_i   N_{c_i}Q_{em,i}Q_{D_i}= Tr ~Q_{em}Q_D=0\,,
\end{equation}
is satisfied, as may likely be expected in the case in a full UV-complete model (which in such a setup, is satisfied then separately for both the $T_{3L}$ and $Y/2$ SM diagonal generators), then 
$\epsilon$ becomes a finite and, in principle, calculable quantity. 
Interestingly, when both the DM and the DP are sub-GeV in mass, one observes that the magnitude of $\epsilon$ is constrained by numerous experiments as well as the value of the relic density 
to lie very roughly in the $\epsilon \sim 10^{-(3-4)}$ range, a value that we might have expected since it originates from a loop process. Importantly, in such a setup, for either ($i$) $p-$wave annihilating 
DM (\eg, it being a complex scalar) or ($ii$) for pseudo-Dirac DM with a sufficiently large mass splitting, it is found that the somewhat tight constraints arising from the CMB on DM annihilation 
into electromagnetically interacting final states can be essentially avoided\cite{Planck:2018vyg,Slatyer:2015jla,Liu:2016cnk,Leane:2018kjk} for a significant range of model parameters. 
In addition, this same set of parameter choices also allows one to escape other significant constraints in this DM mass range arising from, \eg, the galactic X-ray flux\cite{Koechler:2023ual} and the 
511-keV line data\cite{DelaTorreLuque:2023cef}. 

Now in a purely bottom-up approach, one might not expect Eq.(2) to hold in general except by construction, without it being due to any particular symmetry or relationship between the $Q_D$ 
and $Q_{em}$ (or more generally, $T_{3L}$ and $Y$) charge generators. In such a case, this condition is more that likely insufficient to insure that $\epsilon$ will remain finite and calculable 
at higher orders in perturbation theory. However, as noted, our eventual goal is the construction of a fully UV-complete,  `GUT -like' model wherein such a symmetry can naturally exist and is 
typically found to be quite common in such setups. For example, consider the familiar breaking of $E_6 \to SU(5)\times U(1)_\chi \times U(1)_\psi$ where the SM is embedded in $SU(5)$ and now let 
$Q'$ be an arbitrary linear combination of the $U(1)_\chi$ and $U(1)_\psi$ generators. Then it is easily shown, by employing the well-known orthogonality of the $E_6$ generators, that for 
any complete representation, $R$, one finds that
\begin{equation}
Tr_R   ~(T_{3L},Y) Q'=Tr_R~Q_{em}Q'=0\,,
\end{equation}
which is just one example of Eq.(2) above. When such a condition holds naturally as a result of the UV-complete, `GUT-level' gauge symmetry which is then only spontaneously broken as in this 
example, it maintains the finiteness and calculability of $\epsilon$ through to higher orders. We will, however, have no need of making any direct use of this particular result in the analysis that 
follows below. 

Now once we independently posit the existence of PM, other loop diagrams can also be contemplated which may induce a different type of interaction between the SM fields and DM, particularly as 
it is likely that the $U(1)_D$ gauge group must be embedded into some larger (at least partially) non-abelian structure before a very large mass scale 
is reached\cite{Davoudiasl:2015hxa,Reilly:2023frg,Rizzo:2022qan,Rizzo:2022lpm}. The reason for this, and as we'll briefly review below, is that once we know the particle content of the model 
(be it either ($i$) or ($ii$) above) at the sub-GeV scale, $M_L$, we can use the renormalization group equations to run the $U(1)_D$ gauge coupling, or more precisely $\alpha_D=g_D^2/4\pi$, to larger 
scales, $M_U${\footnote {Recall that SM fields will play no role here as they all have $Q_D=0$}}. 
Generally one finds that, as $U(1)_D$ is not asymptotically free, for a significant span of low scale values, $\alpha_D$ becomes non-perturbative (or even runs into a Landau pole) when 
mass scales in the range $\sim$ a few to 10's of TeV are approached and this can only be remedied by an embedding of $U(1)_D$ into this larger non-abelian group, $G$, which is asymptotically 
free {\it before} this scale is reached. Some possibilities for $G$ have been discussed in earlier work\cite{Rizzo:2018vlb,Rueter:2019wdf,Kim:2019oyh,Rueter:2020qhf,Wojcik:2020wgm,Rizzo:2021lob,Rizzo:2022qan,Wojcik:2022rtk,Rizzo:2022jti,Rizzo:2022lpm,Wojcik:2022woa,Carvunis:2022yur,Verma:2022nyd,Rizzo:2023qbj,Wojcik:2023ggt,Rizzo:2023kvy}; a general feature of these analyses is that one usually finds that at least some of the particles of the SM will lie in common 
representations of $G$ together with some of the PM fields. This means that there will exist heavy gauge bosons in $G$ which will link these two sets of fields and can then yield a 
loop-induced, three-point coupling of the SM with the DP, $V$. As shown in earlier work\cite{Rizzo:2021lob} (and see also \cite{Barducci:2021egn}), hereafter referred to as I, this interaction is 
similar to that induced between the electrically neutral SM neutrino or DM{\footnote {See, \eg, Refs.\cite{Hambye:2021xvd,Sharma:2023jdo}}}  
with the photon and can be expressed at low energies in the form of (now dark) dipole moments and other form factors for the SM fermions. While the nature of this interaction is clearly quite different than that generated via KM, it was found that this scenario could just as (or more) easily satisfy all of the existing low energy 
experimental constraints over a respectable range of parameters similar to those in the KM 
picture. We note, however, for this approach to be viable the PM fields must be somewhat heavy vector-like fermions\cite{CarcamoHernandez:2023wzf}.
Although the physics of the KM and that of the `dark moments' portals would appear to be quite different below $M_U$ (apart from the running of $\alpha_D$ which is unaffected by this choice), 
once $M_U$ is actually reached, both scenarios would share many of their common elements: PM fields, new heavy gauge bosons and the heavy dark Higgs fields necessary to break the group 
$G$ down to $U(1)_D$. In both cases, one of the simplest possibilities for $G$ is the SM-like structure $G=SU(2)_I\times U(1)_{Y_I}$, as was discussed in 
Ref.\cite{Rueter:2019wdf}\footnote{See also Ref.\cite{Bauer:2022nwt}} and which was motivated by $E_6$-type gauge models\cite{Hewett:1988xc} 
although somewhat more complex scenarios are easily constructed, \eg, in Refs.\cite{Wojcik:2020wgm,Rizzo:2023qbj,Rizzo:2023kvy}. 

Unlike in the KM scenario where the strength of the SM-DM interaction, as measured by $\epsilon$, depends only upon the ratios of the various PM masses, in the dark moment approach, the 
overall scale of the masses of these new fields is also relevant and for this setup to be phenomenologically successful these states cannot lie too far above the TeV scale. This further 
strengthens the arguments made above, based on the running of $\alpha_D$, about the scale $M_U$ not being allowed to be very high. However, 
LHC searches have already been shown to lead to some significant constraints on these various types of new particles that need to live near the scale $M_U\simeq$ a few TeV. Given this, one might 
ask what would happen if $M_U$ were only a bit larger than the current bounds, by no more than a factor of 2-3, so that the these new states would all lie beyond the reach of the HL-LHC 
or have production cross sections which are too small to be observed? Clearly, until much higher energy machines, such as FCC-hh or multi-TeV muon colliders, are constructed in the relatively 
distant future, we will have, 
at best, only an indirect window into the physics at the $M_U$ scale for {\it either} the KM or dark moments portal pictures. If either (or some combination of) these models were 
to be realized, what would we be able to learn about the nature of this higher scale physics from the set of  indirect measurements made at lower energies? In addition to the numerous (very) 
low energy measurements that are or soon will be available below the $\sqrt s \simeq 10$ GeV scale of Belle II, data from $e^+e^-$ colliders in the $\sqrt s=M_Z-1$ TeV intermediate energy range, but still operating below the scale $M_U$, will at some point also come into play, and may provide the needed information for us. In order to address this question we will make use of a simplified, 
toy model version of the $G=SU(2)_I\times U(1)_{Y_I}$ scenario discussed above which was first introduced in Ref.\cite{Rizzo:2021lob} wherein the SM electron shares a representation of $G$ 
with (at least) one PM field. Within this simple model will consider both the KM and dark moment setups and ask what if anything we can deduce about PM scale physics via these indirect 
$e^+e^-$ measurements. 

The outline of this paper is as follows:  Following the present Introduction, in Section 2,  we will return to the toy model presented in I, generalizing the basic setup there to allow for higher energy 
interactions of the SM fermions with the DP, and discuss how it can be used to think about the SM interactions with DM generated by either the KM or dark moment model structures. 
Generally, this will mean that in the dark moment case we must (at least) include operators/terms of the next higher dimension than those employed in I that will grow more strongly with $q^2$. We will 
then return to the issue of a likely upper bound on the value of $M_U$, which is common to both of these setups, for the two differing assumptions made above about the nature of the DM briefly mentioned 
above.  In Section 3, we explicitly extend the toy model predictions for the dark moment interaction to the first two next to leading order terms in $q^2$ as would be relevant for measurements made at 
intermediate energies, $\sqrt s=M_Z-1$ TeV, but still below the on-shell PM production scale which we assume to be $O(M_U)\sim$ several TeV. In Section 4, we provide an overview and background 
discussion for the testing of both of these model frameworks at future $e^+e^-$ colliders running in this intermediate energy range.  Section 5 explores the issues associated with the precision 
measurement of the radiative $e^+e^-\to$ DM process at intermediate energies at colliders as a means to indirectly probe $M_U$-scale physics in both toy model setups. Here we show that the 
situation in the KM framework seems quite hopeless but at least some portions of the dark moments toy model parameter space may be more amenable to these efforts. In Section 6, we examine 
how dark moments may alter the production of purely SM final state processes, \eg, $e^+e^-\to \mu^+\mu^-$ as well as Bhabha scattering, and their usefulness in exploring $M_U$ scale physics within 
the toy model framework. As was the case for DM pair production,  it again seems likely that a significant range of the toy model parameters will allow us to extract information about $M_U$ 
scale physics here as well.  A summary and our conclusions can be found in Section 7.

%-------------------------------- DOCUMENT: SECTIONS -----------------------------------------%

\section{Background and Overview of General Setup}

In order to indirectly probe what new physics may be active at the PM scale we must first resolve the issue of how the SM-DM interaction is generated.  As was noted in the Introduction, 
we will exclusively consider one of two extreme portal scenarios wherein the low energy interaction between the Dark Sector and the SM are in either case generated by loops of PM fields.  
For concreteness in the following discussion these are specifically identified as vector-like fermions (VLF) carrying charges from both sectors and which live in a more UV- complete structure at a 
larger scale, $M_U \sim$ TeV scale or above. As was discussed in I, both of these setups can lead to thermal DM at the sub-GeV mass scale for a respectable range of parameters while also avoiding 
other existing constraints. We will then attempt to address the question raised in the Introduction: ``What can indirect measurements, made at intermediate energies below the scale 
$M_U$, which are currently or `soon' to be accessible, tell us about the physics at/above it?".  Specifically, in terms of the toy model employed here, we'd like to know what such measurements can 
tell us about the presence of the new fields that exist in this model, \eg, their masses and interactions.

The physics in the KM scenario is very familiar but that of dark moment case is less so; here, we review this setup quite generally and then focus on the framework of the toy model introduced in I 
as will be needed for our further discussion{\footnote {Some of the phenomenological difficulties of the original toy model are discussed in I. It can be made more realistic\cite{Rizzo:2023qbj} but only 
at the price of introducing many more fields and additional parameters thus making the essential physics far less transparent, especially for many aspects of the important aspects of the 
present discussion.}}.  To begin, it is well-known that the interaction of a SM Dirac fermion having a vanishing dark charge at tree-level with a single DP, $V$, can, in analogy with the ordinary 
photon, be described by the three-point effective vertex 
\begin{equation}
{\cal L} = i\bar f(p')\Gamma_\mu f(p) V^\mu (q)\,,  
\end{equation}
with $q=p-p'$ and where for the case of on-shell external SM fermions, $f$, we may write in all generality (here very loosely employing the notation of Ref.\cite{Chu:2018qrm}) that 
\begin{equation}
\Gamma^{\rm {on}}_\mu = e\epsilon Q_f \gamma_\mu+\frac{i\sigma_{\mu\nu} q^\nu}{\Lambda_2}\big[M(q^2)+iE(q^2)\gamma_5\big]+\frac{(q^2\gamma_\mu-q_\mu \qslash)}{\Lambda_1^2}\big[C(q^2)-A(q^2)\gamma_5 \big]\,,
\end{equation}
where we have explicitly extracted inverse powers of the dimensionful overall scale factors, $\Lambda_{1,2}$, and where $M,E,C$ and $A$ are now dimensionless, $q^2$-dependent, form 
factors. Here, $Q_f$ is just the 
usual SM electric charge of the fermion $f$ and, as above, $g_D$ is the $U(1)_D$ gauge coupling. In the limit of CP-conservation and also of parity-conserving SM-PM interactions, as will be 
the case for the toy model introduced in I and that will concern us in the discussion below, we can rewrite this three-point vertex in a somewhat more familiar notation simply as, 
\begin{equation}
\Gamma^{\rm {on}}_\mu = e\epsilon Q_f \gamma_\mu+\frac{ig_D\sigma_{\mu\nu} q^\nu}{\Lambda_2}F_2(q^2)+g_D\frac{(q^2\gamma_\mu-q_\mu \qslash)}{\Lambda_1^2}\tilde F_1(q^2)\,,
\end{equation}
where $F_2$ is the usual magnetic dipole moment form factor while $\tilde F_1$ is a truncated charge-like form factor defined such that last term in the above expression vanishes as $q^2\to 0$ for 
a DP coupling to a conserved current, as will always be the case below. Note that with this assumed structure, the leading term in both $F_2$ and $\tilde F_1$ for small $q^2$ is just 
a constant followed by a power series in $q^2/m^2$ where $m$ is some large scale of order that of the PM masses. This will be seen explicitly in the toy model below.

\begin{figure}[htbp]
\centerline{\includegraphics[width=2.6in]{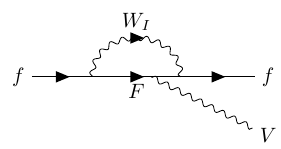}
\hspace{1.0cm}
\vspace {1.5cm}
\includegraphics[width=2.6in]{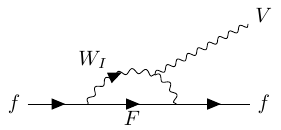}}
\vspace*{-0.40cm}
\caption{One-loop vertex diagrams that can generate the dark moment type interaction terms in the toy model as discussed in the text from I. There are no graphs where the dark photon, $V$, is 
emitted from the external lines as the SM fermion, $f$, is assumed to be neutral under $U(1)_D$, \ie, has $Q_D=0$.}
\label{fig1}
\end{figure}

In the first, quite familiar, portal scenario that we consider, the interaction of the DP with the SM fields is generated by the PM fields in the vacuum polarization-like one-loop diagrams 
discussed previously which leads to the usual KM of the DP with the SM photon; this type of dimension-4 coupling is described by the familiar first term in the interaction above with the loop suppression 
found in the small parameter $\epsilon$. The second, somewhat less familiar possibility,  as discussed in I\cite{Rizzo:2021lob}, is that KM is for some reason suppressed and the $\bar ffV$ 
three-point function is now {\it directly} induced by the one-loop vertex diagrams such as those shown in Fig.~\ref{fig1} from I. In this case, this interaction corresponds to the second and third terms 
in the three-point function decomposition above and we will refer to this setup below as the Dark Moment (DMom) 
portal scenario to contrast this with ordinary KM. The leading contribution to this interaction arising from these terms in the $q^2\to 0$ limit, as is relevant al lower energies, say 
$\sqrt s \leq 10$ GeV, was the subject of 
the previous work in I \cite{Rizzo:2021lob}. We note that while we will take an either/or approach to these two portals in what follows, it is not unlikely that both of them may be active simultaneously. We 
also recall from these earlier discussions that whereas other charged SM fields, such as the $W^\pm$, will also interact with the DP through the same type of  $\epsilon$-suppressed coupling in the 
KM scenario,  $W^\pm$ will {\it not} have one-loop induced DMom-type couplings in the present realization of the DMom setup even though such coupling structures are easily written down in 
a fashion analogous to the familiar anomalous gauge couplings. This is due to the fact that the PM fields themselves are here, by assumption in this toy model, fully vector-like as far as all of their 
gauge interactions are concerned and so a $W^+W^-V$ coupling induced by a PM fermion loop is forbidden at this level by the usual charge conjugation invariance arguments.

In our previous consideration of the DMom scenario, we limited our discussion to energies at or below that of the BaBar and Belle II B-factories, $\sqrt s \sim 10$ GeV, and as such only the 
leading term in $F_2$ (\ie, essentially a scaled $\Lambda_2$) was relevant to the discussion as terms arising from $\tilde F_1$ (as well as the higher order terms in the $q^2$ expansion of $F_2$) 
are necessarily highly suppressed by powers of large mass scales. Thus, apart from the sub-GeV particle masses themselves, only the values of $\Lambda_2$ and $\alpha_D=g_D^2/4\pi$ 
were phenomenologically relevant for making any predictions in our previous study. As we go to higher, intermediate energies (but still remaining at scales below $M_U$), this will no longer 
be the case and, at the very least, we might expect that the lowest sub-leading term in $F_2$ as well as the leading term proportional to $\tilde F_1$, \ie, the value of $\Lambda_1$, will now 
also become of some numerical importance.  We assume that this will be the case in the discussion that follows and then use these new terms to potentially access $M_U$ scale physics.

\begin{figure}[htbp]
\centerline{\includegraphics[width=5.0in,angle=0]{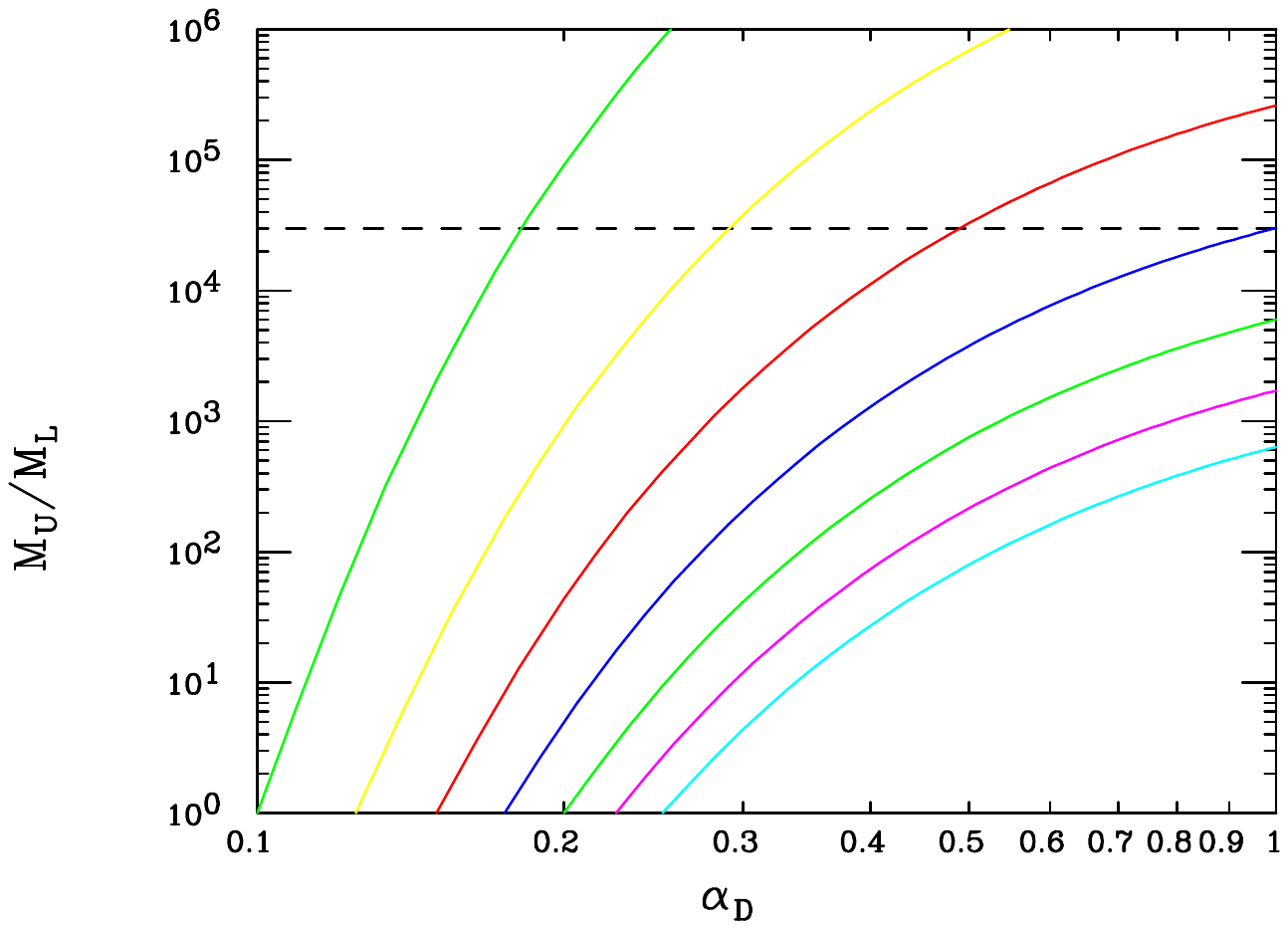}}
\vspace*{-0.8cm}
\centerline{\includegraphics[width=5.0in,angle=0]{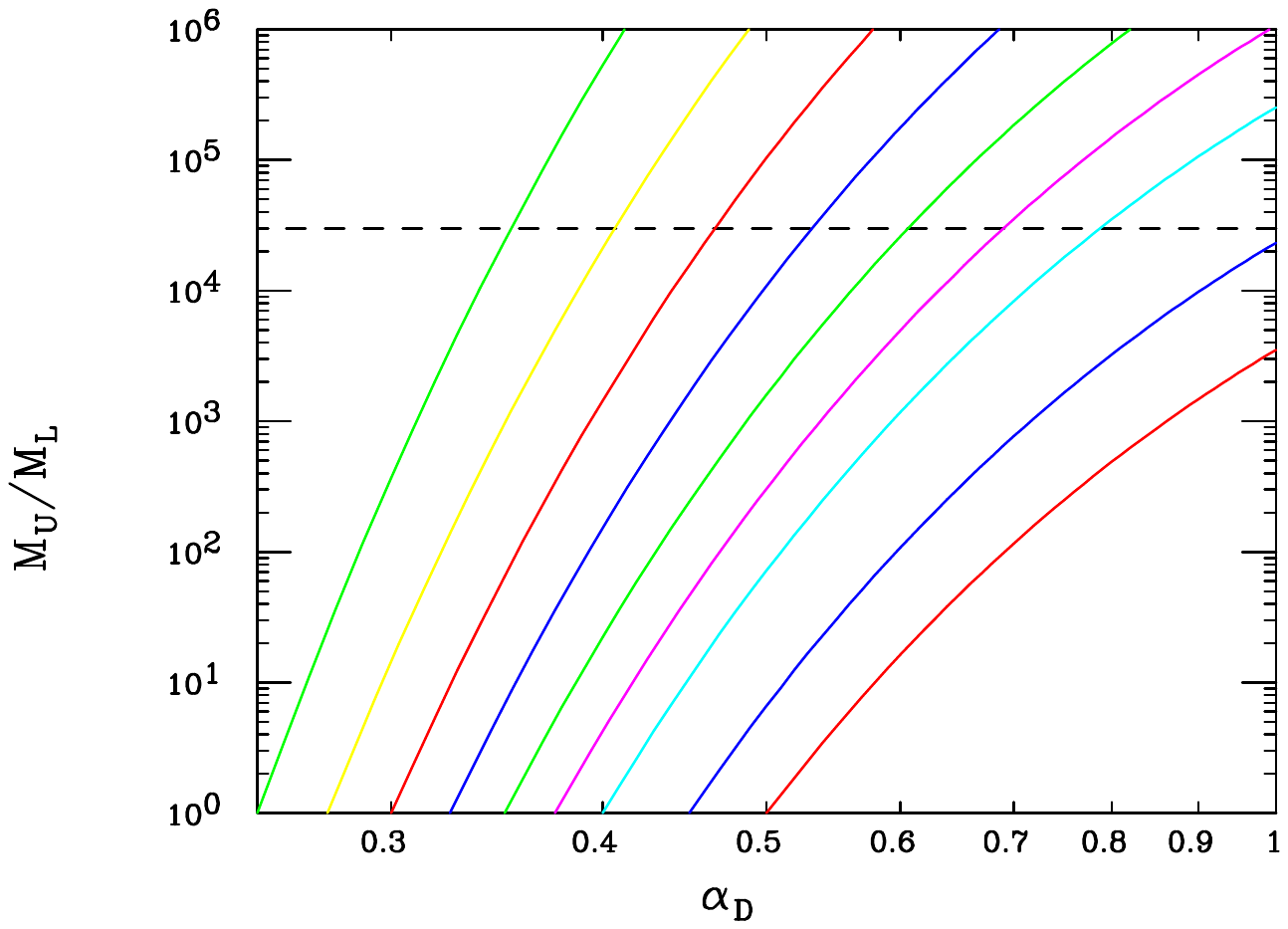}}
\vspace*{-1.3cm}
\caption{The value of $M_U/M_L$ as a funtion of $\alpha_D$ employing 3-loop running in the MS-bar scheme for either the assumption of (Top) pseudo-Dirac or (Bottom) scalar DM. The curves 
from right to left in the Top panel assume that $\alpha_D(M_L)=0.250,0.225,...,0.100$, respectively, whereas the corresponding curves in the Bottom panel are for the values 
$\alpha_D(M_L)=0.50,0.45,0.40,0.375,0.35,..,0.25$, respectively. The dashed line, to guide the eye in both panels, corresponds to $M_U=3$ TeV when $M_L=0.1$ GeV. Potential contributions from 
the DH quartic and from the DM-DH Yukawa coupling are ignored here.}
\label{fig2}
\end{figure}

Before this discussion, we need to justify our expectation that for much of the parameter space, $M_U$ is likely not too far away from the TeV mass scale so that indirect probes are in principle 
possible. To begin our analysis and recalling earlier studies\cite{Davoudiasl:2015hxa,Reilly:2023frg,Rizzo:2022qan,Rizzo:2022lpm}, we notice that both of the KM and DMom portal scenarios 
have a single common feature at scales significantly below 
$M_U$.  Based on the low energy, $\lsim 1$ GeV, particle content of the model and the fact that SM fields do not carry a $U(1)_D$ charge, one can examine the running of the corresponding 
gauge coupling and estimate a bound on the mass scale $M_U$ by asking where we might expect the value of $\alpha_D$ to become non-perturbative or even have 
a Landau pole. We should then expect that 
before this can happen $U(1)_D$ must be embedded into a larger, non-abelian gauge group.  One advantage of probing the RGE running of $\alpha_D$ is that it only depends upon the low 
energy content of the model 
and {\it not} upon the detailed PM physics that takes place at/near $M_U$ generating either the KM or DMom interaction structures between the DP and the SM that we are examining here. 
This type of analysis has been performed several times in the literature employing various input assumptions but always yielding semi-quantitatively quite 
similar results\cite{Davoudiasl:2015hxa,Reilly:2023frg,Rizzo:2022qan,Rizzo:2022lpm}. Fig.~\ref{fig2}, which generalizes the results from Refs.\cite{Rizzo:2022qan,Rizzo:2022lpm}, shows the 
3-loop running of $\alpha_D$ in the MS-bar scheme in the limit that the contribution to the running from the Dark Higgs (DH) quartic and, in the case of pseudo-Dirac DM, the corresponding 
DM-DH Yukawa coupling, can be ignored{\footnote {Possible terms of $O(\epsilon^2)\lsim 10^{-7}$ have also been neglected in obtaining these results.}}. 
Here, the low mass scale, $M_L$, is that corresponding roughly to the DM, DP and DH masses which are assumed to all be lying in the range 
$\sim 0.1-1$ GeV. From this Figure we see, \eg, that for the case of pseudo-Dirac (scalar) DM, $\alpha_D$ will become non-perturbative at or below $M_U=3$ TeV if $\alpha_D(M_L)$ is greater 
than  $\simeq 0.175$ ($\simeq 0.420$) when we take $M_L=100$ MeV, corresponding to the ratio $M_U/M_L=3\cdot 10^4$. 

Of course the large values of $\alpha_D$ as shown in this Figure are not meant to justify their usage but to show that if they {\it were} to be realized at low scales then the RGE evolution to even larger 
coupling values as the relevant scale increases points to a needed, at least partial, embedding into an non-abelian group (to flip the sign of the $\beta$-function) at energies not too far above those 
currently being studied. However, it is to be noted that this coupling range is observed to have a significant overlap with the values chosen in many phenomenological studies at low energies at/below 
the $\lsim 10$ GeV scale\cite{Alexander:2016aln,Battaglieri:2017aum,Bertone:2018krk,Cooley:2022ufh,Boveia:2022syt,Schuster:2021mlr}. Obviously, for a fixed value of $\epsilon$, since it scales as 
$\sqrt \alpha_D$, increasing $\alpha_D$ would require a somewhat higher degeneracy in the PM masses or a (finer) cancellation amongst the terms contributing to Eq.(1).

\begin{figure}[htbp]
\centerline{\includegraphics[width=5.0in,angle=0]{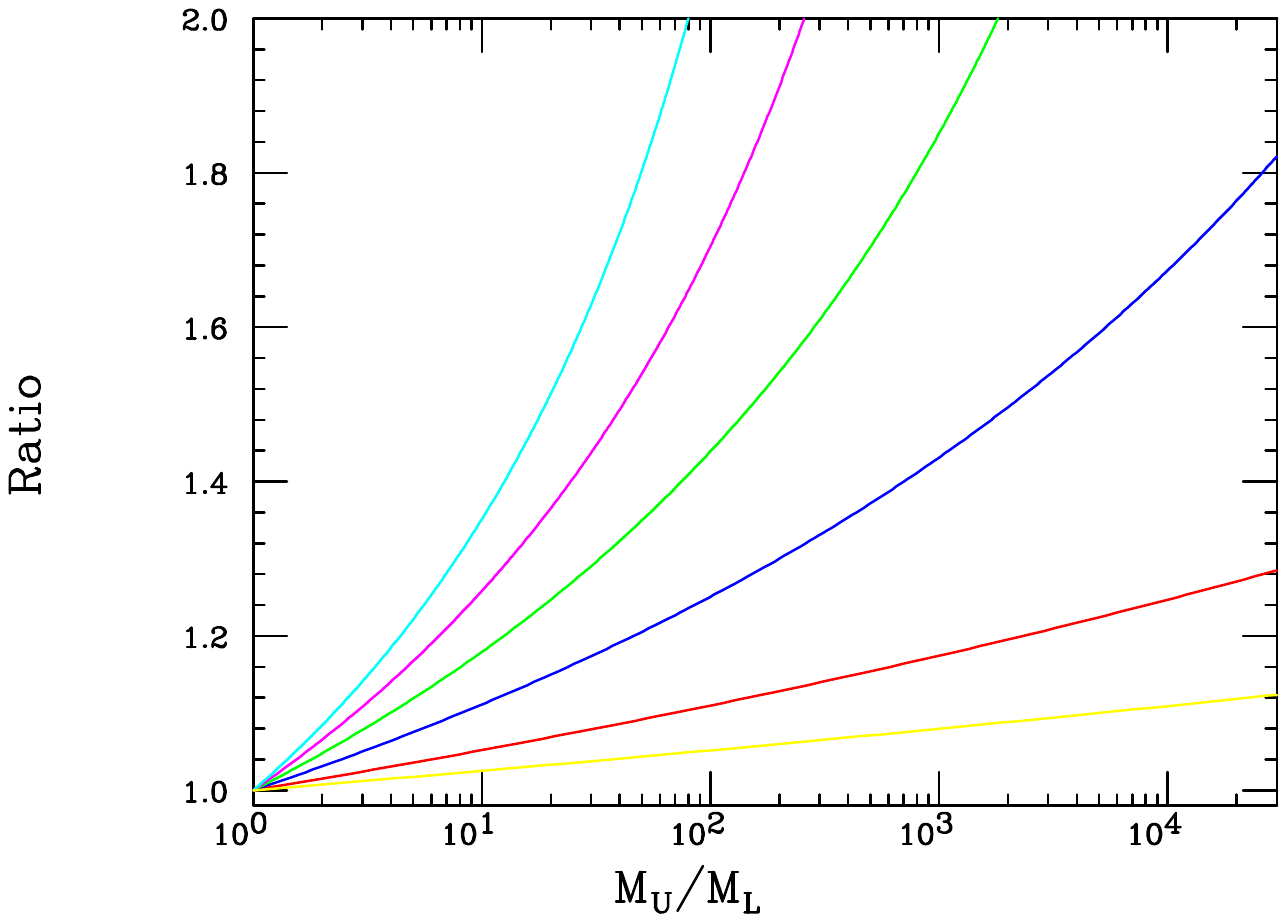}}
\vspace*{-0.8cm}
\centerline{\includegraphics[width=5.0in,angle=0]{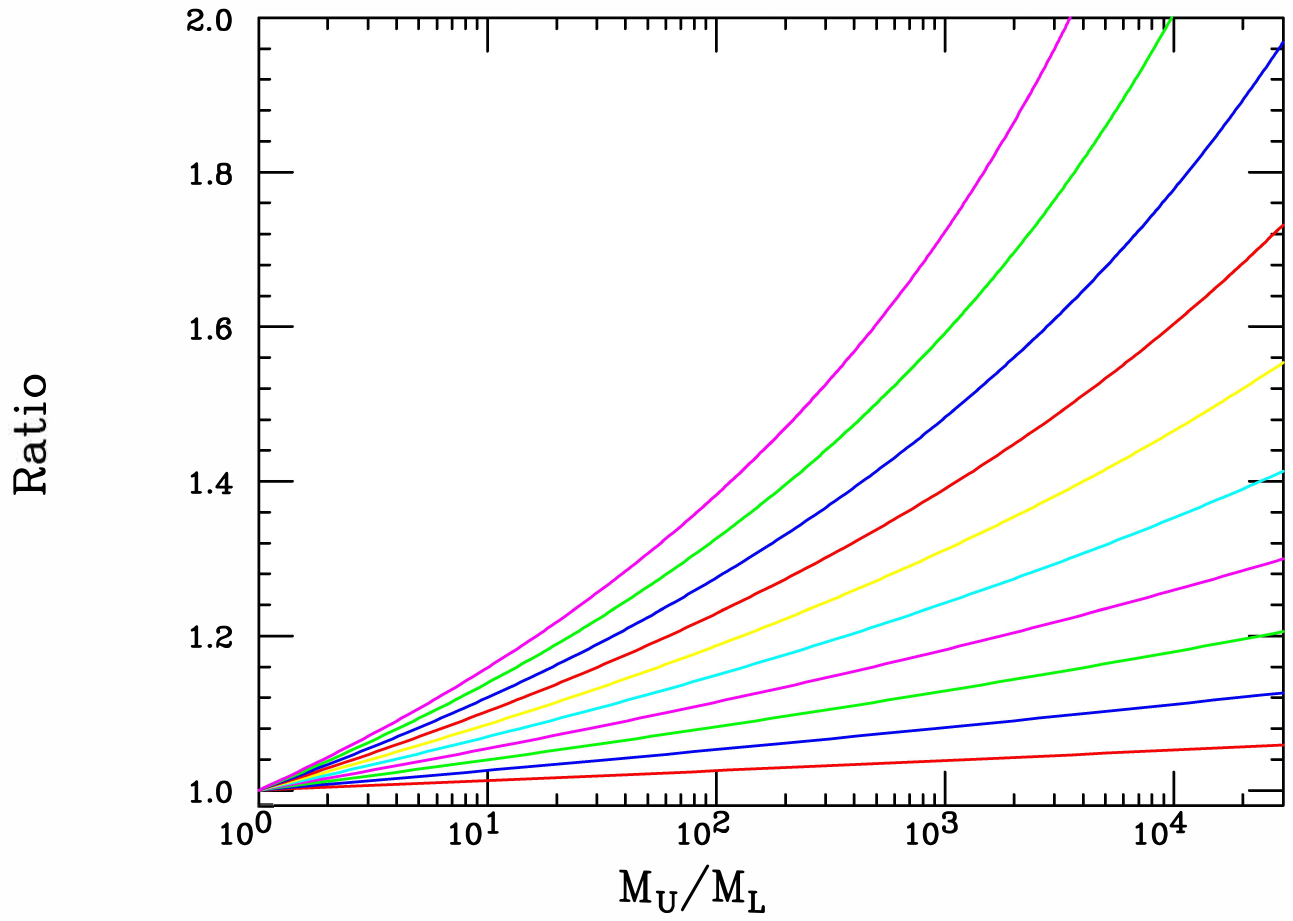}}
\vspace*{-1.3cm}
\caption{The behavior of the ratio $\alpha_D(M_U)/\alpha_D(M_L)$ at the 3-loop level in the MS-bar scheme for the same assumptions as in the previous Figure as a function of $M_U/M_L$ for 
different assumed values of $\alpha_D(M_L)$ for 
the case of (Top) pseudo-Dirac or (Bottom) scalar DM. In the Top panel, from right to left, $\alpha_D(M_L)=0.025, 0.05,0.10,..,0.25$, respectively, whereas in the Bottom panel, it ranges from 
0.05 to 0.50 in steps of 0.05, respectively.}
\label{fig3}
\end{figure}

Under these same set of assumptions, Fig.~\ref{fig3} gives a somewhat different perspective on these same results displaying the ratio of $\alpha_D(M_U)/\alpha_D(M_L)$ as a function 
of the corresponding value of $M_U/M_L$ and which shows the rapidity with which $\alpha_D$ is actually running. This format will be useful below when we discuss the measurements involving  
the running of the $U(1)_D$ gauge coupling relative to its value at $M_L$ as would be determined by, \eg, the value of the DM relic density and low energy experiments. From this Figure 
we can see that this RGE running becomes more rapid as $M_U/M_L$ increases and, \eg, going from $M_L=0.1$ GeV up to the 0.25-1 TeV scale of possible future $e^+e^-$ colliders, the ratio of 
coupling strengths can be quite large, possibly increasing even by up to a factor $\sim 2$ over this range. This rapid change in coupling strength with energy will impact our discussion below.

\section{Higher Order Terms in DMom - Revisiting the Toy Model}

To address some of the issues raised in the previous Section about the DMom scenario, we must return and remind ourselves of some of the details of the simple, yet representative, toy 
model that was introduced in I. This scenario was based on the work in Ref.\cite{Rizzo:2021lob} and which was then more fully realized in a rather complex manner in Ref.\cite{Rizzo:2023qbj}. 
The sole purpose of this toy model in I was to obtain typical and suggestive values of the scale parameter $\Lambda_2$ that could then be used to compare with a broad set of low 
energy experimental constrains. Here we need to return to this simple model in order to extend this analysis to, amongst other things, determine the typical size of the next to leading terms in 
the $q^2$ expansion in the $\bar ffV$ interaction. This setup, or a slight generalization thereof, can then be used to probe the UV sensitivity of measurements made at intermediate energies 
below the scale $M_U$ for {\it either} the DMom (or KM) scenario(s).

The basic idea is to imagine that typical SM fermion fields, \eg, the electron, $e$ (with which we will be mostly concerned with below),  
lie in a representation of a larger non-Abelian gauge group into which $U(1)_D$ is embedded at a scale 
$\sim M_U$ and they share this representation with {\it at least} one PM field, \eg, $E$, having the same QCD and electroweak quantum numbers. In the simplest version of our toy model, the usual 
$U(1)_D$ is embedded in a SM-like $SU(2)_I\times U(1)_{Y_I}$ group so that 
together $e,E$ belong to an $SU(2)_I$ doublet with $Q_D=T_{3I}+Y_I/2$ and such that $Q_D(e)=0$ while $Q_D(E)=-1$. In a manner again similar to the SM, there will exist a non-hermitian, 
but electrically neutral, gauge boson, $W_I^{(\dagger)}$ with $Q_D=\pm 1$ linked to the $SU(2)_I$ isospin raising and lowering operators, that connects these two fields with assumed to be vector-like 
couplings, $\sim \frac{g_I}{\sqrt 2} \bar e\gamma_\mu EW_I^\mu+{\rm {h.c.}}${\footnote{Our primary reason for introducing this toy model is to motivate such a coupling between SM and PM 
fermion fields.}}. It is further assumed that $E$ as well as both of the toy model gauge fields, $W_I^{(\dagger)},Z_I$,  (but not the DP) all have masses near the 
scale $\sim M_U$ up to $O(1)$ factors. As has been much discussed, we note that in such a setup, the vector-like PM lepton $E$ will via mixing, by far, dominantly decay as $E\to eV$ and not into the 
SM Higgs or gauge bosons. Recasting the existing 13 TeV LHC searches with $L=139 ~fb^{-1}$ of integrated luminosity, the authors of Ref.\cite{Guedes:2021oqx} have shown that $m_E\gsim 0.90$ 
TeV. At the HL-LHC with $L=3 ~ab^{-1}$, a similar {\it null} search would lead to the corresponding lower bound of $m_E\gsim 1.45$ TeV. 
We will assume that $m_E$ satisfies this bound below so that the HL-LHC is incapable of directly producing PM scale physics at an observable rate, forcing us to rely on indirect signatures.  
Identifying $f=e$ and $F=E$ in Fig.~\ref{fig1}, we can now see how the $\bar eeV$ form factors $\tilde F_1$, $F_2$ are generated and that they  
will be functions of the mass ratio $a=m_E^2/m_{W_I}^2$. We note that by complete analogy with the SM, one observes that now $\alpha_D=s_I^2 \alpha_I$, where $\alpha_I=g_I^2/4\pi$ as usual, 
and the mixing angle $x_I=s_I^2$ is the analog of the usual weak mixing angle, $x_w=\sin^2\theta_w$.  Note that since $x_l \leq 1$, the requirement that $\alpha_I$ be perturbative will likely lead to 
an even stronger upper limit on the value of $M_U$ than those discussed above. If $SU(2)_I\times U(1)_{Y_I}$ is broken only by an $SU(2)_I$ doublet (although this need not 
be the case in general), again in complete analogy to the SM, one has $M_{W_I}=c_IM_{Z_I}${\footnote {It is important to note that in much of the discussion below we will ignore the influence 
of the $Z_I$ gauge boson.}. 
Thus, as far as what concerns 
us here, if $\alpha_D$ were known then the only free parameters in the toy model at the high scale are $m_E, m_{W_I}$ and either $s_I^2$ or $M_{Z_I}$. Of course, we can easily imagine 
slightly more complex toy models wherein the electron shares a larger representation of $G$ with two (or even more) PM fields, \eg, $(E,E')=E_i$, with opposite values of the dark charge 
$Q_D$; we will make use of this possibility as well below when we consider the running of $\epsilon$ in the KM setup as such a model will also render this quantity finite as discussed above.  

As far as the DM itself is concerned, we will assume for simplicity in this setup that it lies in a $|Q_D|=1$, $SU(2)_I$ singlet representation so that it will $\it {not}$ couple in an off-diagonal manner 
to $W_I$ as does, \eg, the electron. This will be of some relevance in our discussions below.  Alternatively, one might imagine that the DM lies in an $SU(2)_I$ doublet together with, necessarily, 
a $Q_D=0$,  SM singlet state which may lead to some interesting model building, especially in the fermionic case, as will be the subject of later work. We will not, however, pursue this significantly 
more complex possibility in the present analysis.

It is important to note that one of the reasons we refer to this setup as a `toy' model is that we are making use of its structure only to motivate the possible magnitude and parameter dependence of the 
three-point functions arising from these graphs while we are simultaneously ignoring any of its other possible implications, \eg, the effects of of $Z_I$ exchange or other potential loop-level effects. 
Of course in the case where the 
electron occupies an $SU(2)_I$ triplet with the $E,E'$ PM fields, it does not couple to the $Z_I$ gauge boson as it has both $T_{3I},Q_D=0$. In any case, we should not take these toy 
models {\it too} seriously in our discussion. 

Given these assumptions, we can proceed to extract the specific interaction pieces that we need in our analysis by employing several sets of results from the existing literature. We begin
by considering the dipole moment-like term proportional to the form factor $F_2(q^2)$; noting that $Q_D(W_I)+Q_D(E)=0$, we find that we can express this term as 
\begin{equation}
\frac{F_2(q^2)}{\Lambda_2}=Q_W~\frac{\alpha_D}{16\pi s_I^2 m_E}~\Big(I_1(q^2)+I_2(q^2)\Big)\,,
\end{equation}
where $Q_W=Q_D(W_I)$ which we can define to be unity, and where $I_{1,2}(q^2)$ are parameter integrals arising from the left ($V$ emission from the fermion line) and right ($V$ emission 
from the gauge boson line) diagrams in Fig.~\ref{fig1}, respectively, and are given explicitly by 
\begin{equation}
I_i(q^2)=a~\int_0^1~dt~ \frac{N_i}{\sqrt{QA_i}}~\tanh^{-1} \Big( \sqrt {\frac{Q}{A_i}}~ t\Big)\,,
\end{equation}
with $a=m_E^2/m_{W_I}^2$, $Q=q^2/4m_{W_I}^2<1$ by assumption and where we have defined 
\begin{equation}
A_1=1-t+at,~~A_2=t+a(1-t);~~N_1=1-t+at/4,~~N_2=t+a(1-t)/4\,.
\end{equation}
We can easily expand this expression in powers of $Q$ as
\begin{equation}
I_i(q^2)=a~\int_0^1~dt~ \Bigg[ \frac{tN_i}{A_i}+Q~\frac{t^3N_i}{3A_i^2}+...\Bigg]\,,
\end{equation}
since we are only interested in the leading correction beyond the familiar $q^2=0$ result and from which analytical results can easily be obtained. From this we see that we can freely define the combination
\begin{equation}
\Lambda_2=\frac{16\pi s_I^2 m_E}{\alpha_D} \simeq 116.4 ~\Big(\frac{s_I^2}{x_w}\Big)~\Big(\frac{m_E}{\rm {1~TeV}}\Big)~ \Big(\frac{0.1}{\alpha_D}\Big)~ {\rm {TeV}}\,,
\end{equation}
which is roughly of the anticipated magnitude from I{\footnote{In making numerical estimates in what follows, we will always assume for simplicity that $s_I^2=x_w$}}, 
and by performing the above integrals and summing we can now write the two leading terms in the $q^2$ expansion as 
\begin{equation}
F_2(q^2)=f_{20}(a)+\frac{q^2}{12m_E^2}f_{21}(a)+O(q^4)\,,
\end{equation}
now neglecting all of the suppressed higher order terms in $q^2$. 
The leading term, $f_{20}$, produces the previously introduced dark dipole moment and has a well-known functional form as given in, \eg, 
Refs.\cite{Leveille:1977rc,Yu:2021suw,Bolanos-Carrera:2023ppu}, while 
$f_{21}$ produces the necessary leading order correction term in $q^2$ we are seeking. It is important to notice the potentially small pre-factor in front of $f_{21}$ as it will be of some numerical 
relevance in the discussion that follows. Explicitly, one finds that 
\begin{equation}
f_{20}(a)=a~\frac{(t_1+t_2)}{(a-1)^3},~~~f_{21}(a)=a^2~\frac{(t_3+t_4+t_5)}{(a-1)^5}\,,
\end{equation}
where the $t_i$ are given by the expressions
\begin{eqnarray}
&~~~~&t_1=(a-1)(a^2+a+4)-6a~log(a)\nonumber\\
&~~~~&t_2=(a-1)(a^2-11a+4)+6a^2~log(a)\nonumber\\
&~~~~&t_3=6(a-1)(a^3-1)\nonumber\\
&~~~~&t_4=-3(a-1)^2(a^2-9a+4)\nonumber\\
&~~~~&t_5=\big[2a^3(a-13)+8(2a+1)\big]~log(a)\,.
 \end{eqnarray} 

In a similar fashion we can also determine the leading piece of the term proportional to $\tilde F_1$ since we expect that the higher order terms can be safely neglected and as we will find below. 
This calculation essentially mirrors that for the neutrino charge radius (although the relevant fermions are neutral with respect to different gauge groups) as is given in, \eg, 
Ref.\cite{Bolanos-Carrera:2023ppu}. We find that
\begin{equation}
\frac{\tilde F_1}{\Lambda_1^2}=Q_W~\frac{\alpha_D}{144\pi s_I^2 m_E^2}~f_1(a)+O(q^2)\,,
\end{equation}
now employing only the term with $q^2=0$ so that we can make a convenient normalization choice for $\Lambda_1$: 
\begin{equation}
\Lambda_1 =m_E~ \Bigg(\frac{144 \pi s_I^2}{\alpha_D}\Bigg)^{1/2} \simeq 33.4 ~\Bigg[ \Big(\frac{s_I^2}{x_w}\Big)~\Big(\frac{m_E}{\rm {1~TeV}}\Big)^2~ \Big(\frac{0.1}{\alpha_D}\Big) \Bigg]^{1/2}~ {\rm {TeV}}\,,
\end{equation}
which is again of the expected magnitude and, by employing Eqs.(43)-(46) in Ref.\cite{Bolanos-Carrera:2023ppu} and summing, we can obtain a relatively compact expression for $f_1(a)$, apart 
from the overall scaling factor of $\Lambda_1$ above, \ie, 
\begin{equation}
f_1(a)=a~\frac{(t_6+t_7)+\frac{a}{2}(t_8+t_9)}{(a-1)^4}\,,
\end{equation}
where now 
\begin{eqnarray}
&~~~~&t_6=(a-1)(25a^2-29a-2)-6(6a^2-9a+2)~log(a)\nonumber\\
&~~~~&t_7=-(a-1)(43a^2-65a+16)+6a^2(5a-6)~log(a)\nonumber\\
&~~~~&t_8=-(a-1)(7a^2-29a+16)+6(2-3a)~log(a)\nonumber\\
&~~~~&t_9=-(a-1)(11a^2-7a+2)+6a^3~log(a)\,.
 \end{eqnarray} 

Once we are open to possible graphs that may be contributing to dim-8 DM interactions of the DM with the SM fields, one may worry that, \eg, in addition to the 1-loop vertex graphs above, box graphs 
with internal PM and $W_I$ states may also contribute to the interactions discussed below. However, here the assumption that the DM is an $SU(2)_I$ singlet field, as mentioned above, comes 
directly into play since, by this assumption, the $W_I$ cannot connect the DM directly to any other state (the DM being a singlet) as it does for $e$ and $E$ in Fig.~\ref{fig1}. Then, under 
the assumptions of the toy model discussed above, the only manner in which the DM can interact with SM fields is via a DP exchange.  This means that there can be {\it no}, \eg, analogous 1-loop 
box graphs, or any other kind of 1-loop graph in which the DM field connects directly to a loop, that we need to consider as possible contributions to this interaction. 

Fig.~\ref{fig5} displays the $a$-dependence of the two functions $f_1$, $f_{20}$ and the ratio $r=f_{21}/f_{20}$ as given by the various expressions above. Here we see that, \eg, $f_1(a)$ is 
always at least a factor of a few to an order of magnitude larger than $f_{2}(a)$ while the ratio $r(a)$ is always roughly an order of magnitude smaller than $f_{20}(a)$ itself. As we will see, both 
of these numerical results will have important impact in the discussion below. 

\begin{figure}[htbp]
\centerline{\includegraphics[width=5.0in,angle=0]{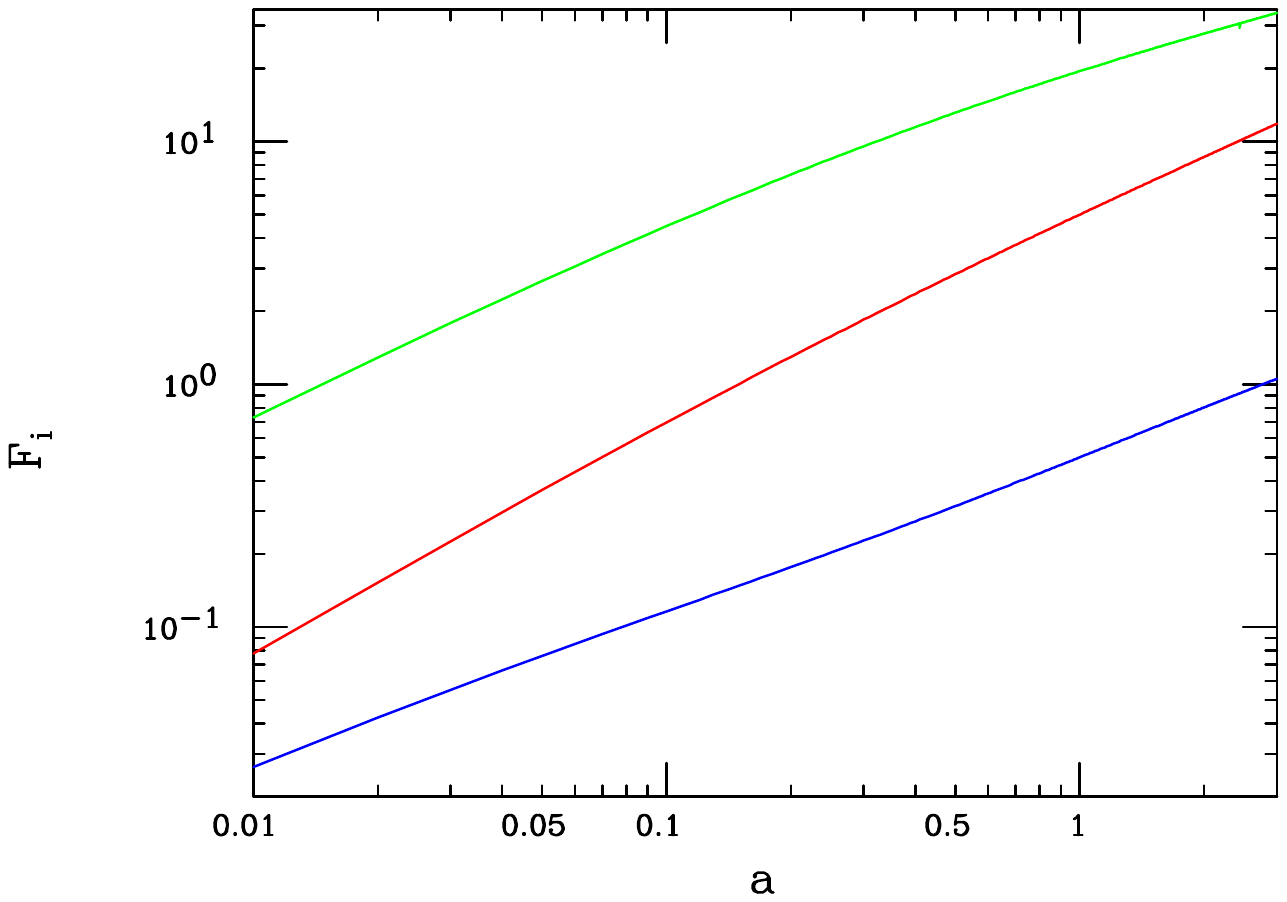}}
\vspace*{-1.3cm}
\caption{The values of the functions $f_1$ (top), $f_{20}$ (middle) and the ratio $r=f_{21}/f_{20}$ (bottom) as functions of the parameter $a=m_E^2/m_{W_I}^2$ for the simple toy model as 
discussed in the text.}
\label{fig5}
\end{figure}

From I, we recall that we already know that $\Lambda_2/(\alpha_D f_{20})\gsim100$ TeV from BaBar null searches for $e^+e^-\to V\gamma$ made at $\sqrt s \simeq 10$ GeV\cite{Lees:2017lec} 
so our parameter choices must respect this constraint. Further, as this $e^+e^-\to V\gamma$ process, with $V$ on-shell, can only probe the $f_{20}$ term of $F_2$, as then $q^2=m_V^2 << s$, 
we need to employ a different process or processes with $V$ off-shell in order to probe the next to leading order terms we are after here. Further, 
it is important to remember that at low energies, as in I, there is essentially only just this one combination of numbers coming from the DMom setup of phenomenological relevance.  With just this one 
number alone we can learn very little about the physics at the $M_U$ scale. However, if we {\it also} know both $r$ as well as the ratio of $\tilde F_1/F_2$ we have a lot more handles on this 
physics. Can we measure these new effects at higher, intermediate energies? This is the question to which we now turn below.

\section{$e^+e^-$ Measurements at Intermediate Energies}

What measurements are possible at energies below $M_U$ that are likely to provide us with information about physics at or above $M_U$? What model parameters are available to be determined? 
In addition to knowing the DP, DM and potentially the dark Higgs masses and other properties\cite{Foguel:2022unm}, that are common to both the KM and DMom setups, we first need a way to 
clearly distinguish these two scenarios to make further progress - but this hopefully may be relatively straightforward in the coming years. Provided a signal for DM is indeed observed, which we will  
assume here, one very important lower energy measurement will, at the very least, cleanly separate the standard KM from the DMom scenarios provided the relevant range of model parameters 
is accessible: the angular distribution of single photon plus missing energy events in $e^+e^-$ collisions arising from the on-shell process $e^+e^- \to V\gamma$, where the $V$ decays invisibly to DM. 
As was explored in I, at BaBar/Belle II energies of order $\sqrt s \sim 10$ GeV, the predicted angular distribution of the photon in the two models is quite different, being essentially flat in the DMom setup 
while scaling like $\sim (1+z^2)/(1-z^2)$, with $z=\cos \theta$, within the usual KM scheme, both in the $m_V^2/s \to 0$ limit. The planned integrated luminosities at Belle II are also likely to eventually 
cover the relevant range of the $\epsilon$ parameter space\cite{Belle-II:2018jsg,Belle-II:2022jyy} and so will discover or discredit either (or both) of these scenarios. Thus, by the time 
that, \eg, $e^+e^-$ Higgs factories will be turning on, we should know if either of these models is in fact realized by nature. We will proceed under the assumption that this has indeed transpired and 
that the choice of KM vs. DMom scenario has been resolved by such measurements; however, we will consider both setups in the discussion here as we don't yet know this outcome. What do we do next?

While the KM and DMom models are designed to work in a similar manner at low energies, we've already seen that they are easily distinguishable by measurements made at $\sqrt s \sim 10$ GeV. 
As we go to the higher energies of interest to us here, these variations will grow even further due to, amongst other things, their significantly differing energy behavior of their predictions below the 
PM scale.

If KM is realized, we need to measure the parameter $\epsilon$ not only at low scales, $\sim M_L$, but also its evolution to higher intermediate ones which is, at least partially, controlled by 
physics at the 
scale $M_U$ about which we wish to learn. Employing the formulae in the Introduction, we will for simplicity consider the case of a pair of leptonic, color-singlet PM vector-like fermion (scalar) fields 
which have $|Q_DQ_{em}|=1$, so that $\epsilon$ is indeed finite, and which have masses $m_1>m_2$. This may be easily arranged in a realistic setup and, \eg, can occur in a modified version 
of our toy model wherein, \eg, in the case of VLF fermions, these two PM fields and the corresponding SM field lie in a triplet of $SU(2)_I$ having $Y_I=0$.  Following Ref.\cite{Guedes:2021oqx},
we will assume, as per the discussion above, a null search result for these states at the HL-LHC so that their masses must lie at or above $\simeq 1.45$ TeV. For finite and calculable $\epsilon$, this is 
an example of the VLF PM fields which can be used to represent the simplest version of new physics at the scale $M_U$ that we want to probe indirectly. 

Within this specific realization of the KM setup, one finds  that 
\begin{equation}
\epsilon_0=\epsilon(q^2=m_L^2\simeq 0)=\frac{g_De}{12(48)\pi^2}~ln \frac{m_1^2}{m_2^2}\,,  
\end{equation}
so that, apart from the known running of the QED coupling, $\alpha_{em}$, the running of $\epsilon$ is then {\it almost} totally controlled at low scales by that of $\alpha_D$, \ie, essentially via the 
square root of the ratio $\alpha_D(M_U)/\alpha_D(M_L)$ shown in Fig.~\ref{fig3}. The remaining {\it additional} running of $\epsilon$ with $q^2$ as arising from {\it high scale physics} as $M_U$ is slowly 
approached from below can also be calculated and is given by 
\begin{equation}
\frac{\epsilon(q^2)}{\epsilon_0}=1+\frac{6}{ln \big(\frac{m_1^2}{m_2^2}\big)}\int_0^1 ~dx~\chi(x)~ln \Big[\frac{1-b_1x(1-x)}{1-b_2x(1-x)}\Big]\,,  
\end{equation}
with $b_i=q^2/m_i^2<4$ and where the simple function $\chi(x)$ is given by the choice one of the two familiar expressions depending upon the nature of the PM:
\begin{equation}
\chi(x)= x(1-x)~~~ {[\rm{fermionic ~PM}}],~~~=x(2x-1)~~~[{\rm{scalar ~PM}}]\,.  
\end{equation}
Given the discussion of the toy model above, here we will limit ourselves to the case of fermionic PM when performing numerical estimates. We again emphasize that this expression above 
encapsulates the running of $\epsilon$ due {\it only} to the high-scale physics we are interested in probing. Below the scale $M_U$ it is only this non-$\alpha_D$  and non-$\alpha_{em}$ contribution 
to the running of $\epsilon$ that allows us to probe PM scale physics. The question we need to address is whether or not any information about this particular piece can be extracted from 
precision measurements once the much more significant energy dependencies of both $\alpha_D$ and $\alpha_{em}$ are accounted for in the data.

As already alluded to, the simplest way to probe the energy dependence of $\epsilon$ above the few-10 GeV energy range is via precision measurements made by employing 
$e^+e^-$ colliders. As noted above, it will be assumed here that the array of existing or `soon'  to be operational experiments (\eg, CMB/relic density, fixed target, collider, direct/indirect 
detection, \etc.) will have been employed to determine the low energy (\ie, below $\sqrt s=10$ GeV) value of $\epsilon=\epsilon_0$ (in the case that KM is realized) as well the masses of the DP, 
DM and the dark Higgs fields.  In such a case, the running of $\alpha_D$ could, in principle, then be calculated from first principles in the same manner as, \eg, the running of $\alpha_{em}$ (away 
from thresholds 
and hadronic resonances) or the way $\alpha_s$ is in QCD. Employing Eq.(19), in Fig.~\ref{fig4}, we show the running of $\epsilon$ due solely to high scale physics, \ie, {\it neglecting} the contributions 
arising from both $\alpha_{em}$ and $\alpha_D$, whose energy dependencies are encapsulated in the relevant RGE's and are due to either the well-known SM fields or to the rather light dark sector 
fields encountered far below 
the scale $M_U$ as was seen in Fig.~\ref{fig3}{\footnote {That same Figure also reminds us that the running of $\alpha_D$ can generally be quite rapid over much of the model parameter space.}}. 
The {\it remaining} energy dependence of $\epsilon$ that we wish to examine is then solely due, in our toy model, to the pair of lepton-like, PM fermion fields, the lightest of which must have a mass 
$\gsim 1.45$ TeV so that they are not observed directly at the HL-LHC as discussed above. 

\begin{figure}[htbp]
\centerline{\includegraphics[width=4.5in,angle=0]{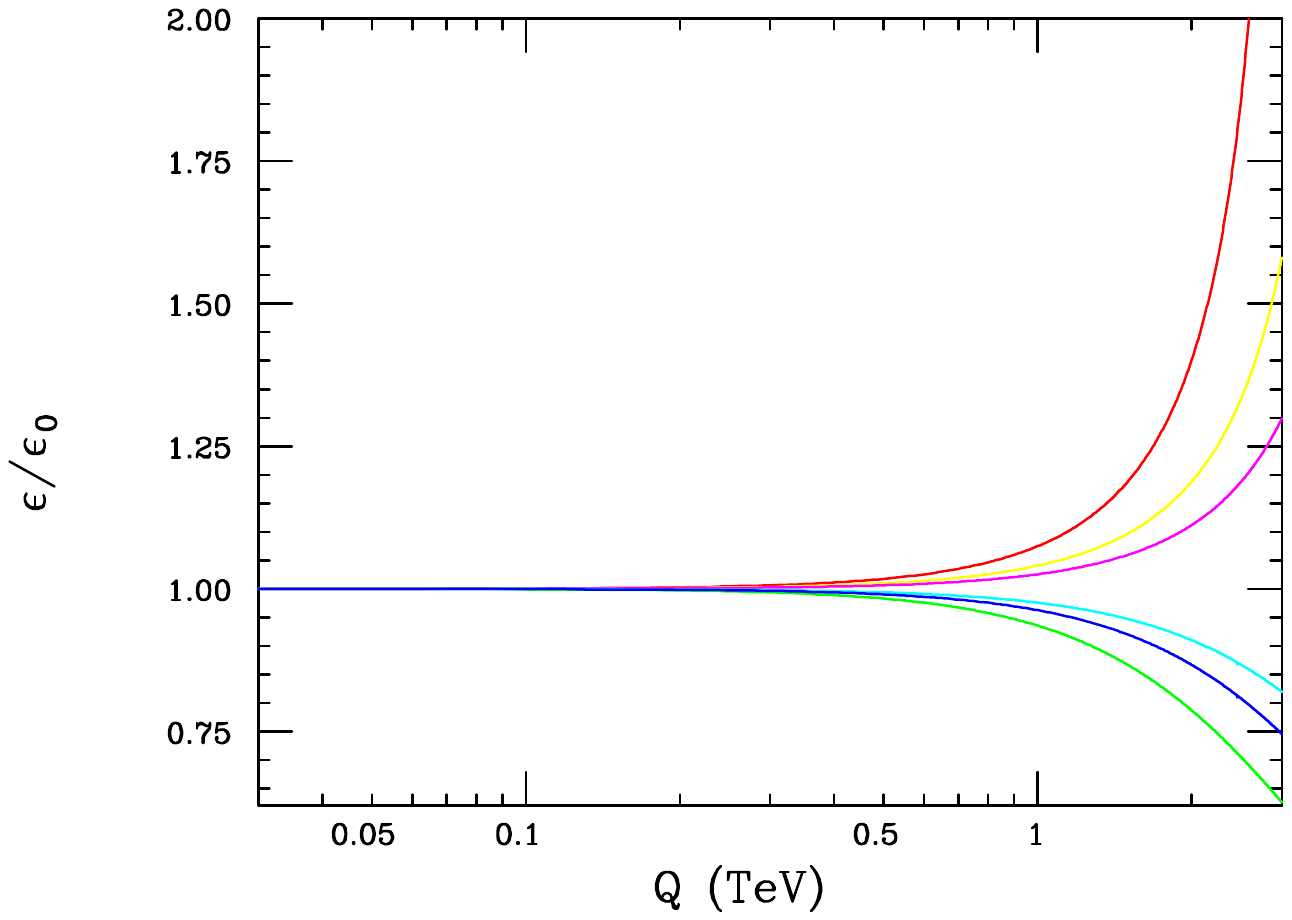}}
\vspace*{-1.8cm}
\centerline{\includegraphics[width=4.5in,angle=0]{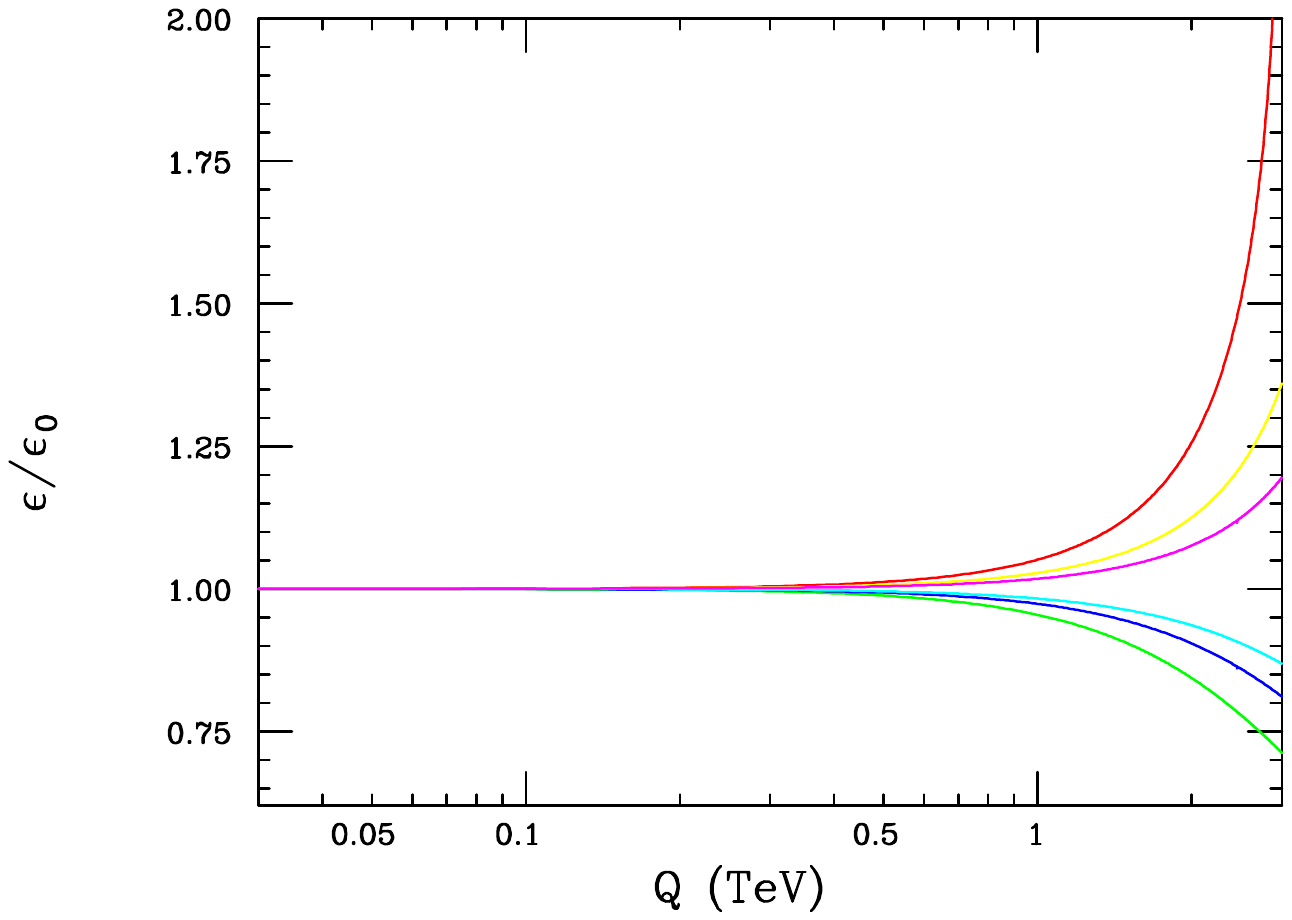}}
\vspace*{-1.8cm}
\centerline{\includegraphics[width=4.5in,angle=0]{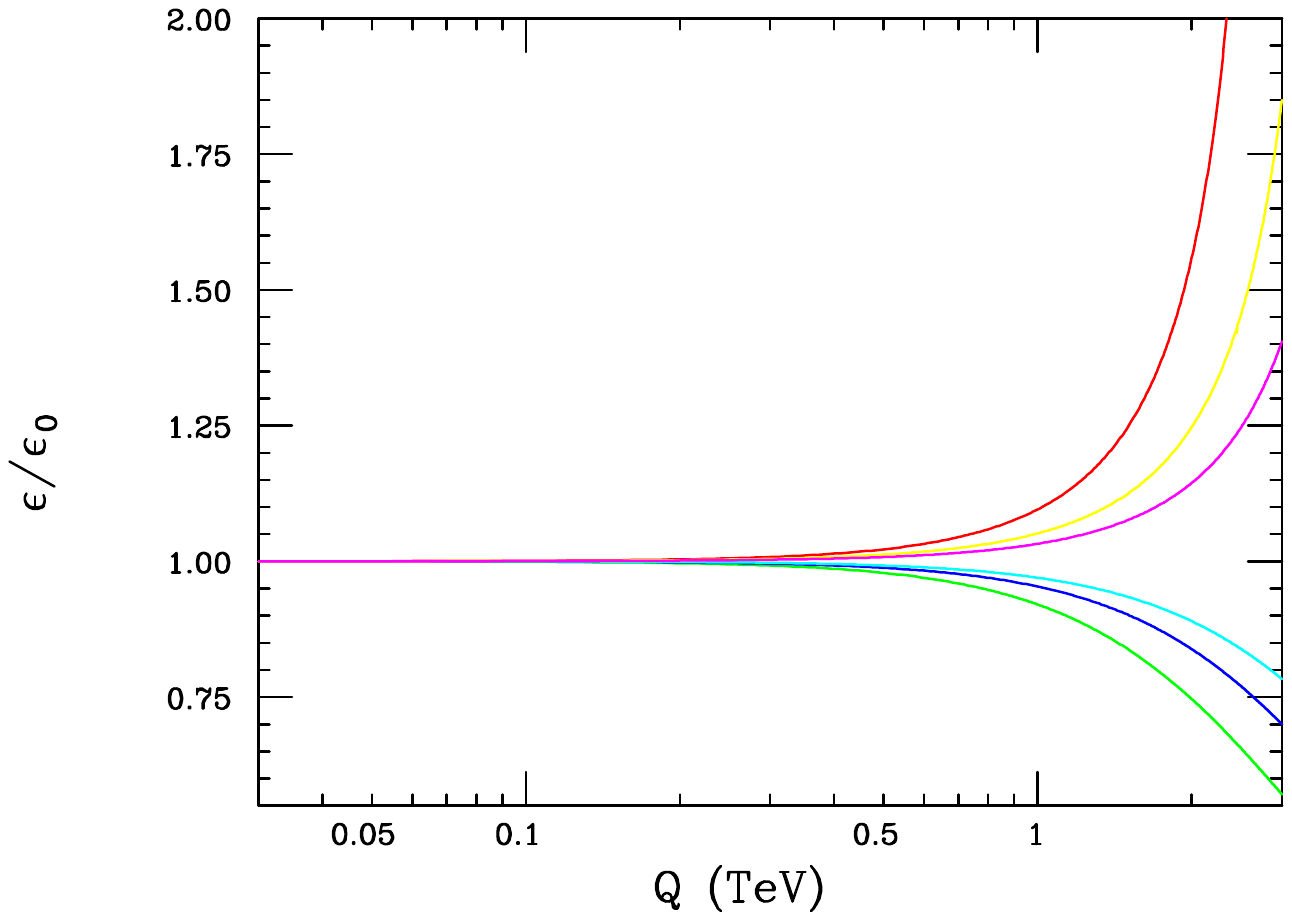}}
\vspace*{-1.1cm}
\caption{The running of the KM mixing parameter, $\epsilon$, due {\it only} to high-scale physics with the running of both $\alpha_{em}$ and $\alpha_D$ turned off, as described in the text, as a function 
of the energy scale $Q$ and normalized to its value at low energies. Here $M_L\sim 0.1-1$ GeV is assumed together with the leptonic VLF PM mass ratios of $m_1/m_2=1.3$(Top), 2.0 (Middle) and 
1.025 (Bottom), respectively. In each panel the upper (lower) set of three curves corresponds to time-like (space-like) values of $Q^2$. In all cases, from top to bottom, the curves 
correspond to the choice $m_2=1.5,2.0,2.5,2.5,2.0$ and 1.5 TeV, respectively.}
\label{fig4}
\end{figure}

An obvious place to start is, in effect, the inverse process to that by which the DM achieves the relic density observed by Planck since we know that such a reaction must occur as discussed in I. 
Here it will be  assumed the heavier dark state produced in association with the DM in the case of the pseudo-Dirac possibility is sufficiently boosted/long-lived so as to appear as missing energy in the 
detector. At collision energies where the electron and DM masses can both be safely neglected, the $e^+e^- \to$ DM pair production cross section in the KM setup is given by the expression 
\begin{equation}
\sigma_e=\frac{(4)\pi ~\alpha_{em}^2 (s)~\epsilon^2(s)}{3s}~\frac{s^2}{(s-m_V^2)^2+\Gamma_V^2m_V^2}\,,  
\end{equation}
when the DM is assumed here to be a scalar (fermion), with the full $s-$dependence as noted in an effective Born approximation. 
In practice, we are specifically concerned with the intermediate energy region far above the DP 
pole, \ie, far from where $V$ can be produced on-shell so that both its mass and width can also be safely neglected. However, we will also simultaneously remain sufficiently below the scale $M_U$ so 
that the VLF PM fields cannot be pair created on-shell. In this same limit, employing the results from above (and now with $q^2=s$), we can decompose the complete $s$-dependence of $\epsilon^2$ as 
being 
\begin{equation}
\epsilon^2(s)=N\alpha_{em}(s)\alpha_D(s){\cal F}(s)\,,
\end{equation}
with $N$ being a (known) number and with the function ${\cal F}=(\epsilon/\epsilon_0)^2$ describing the dynamics embodied in energy dependence of $\epsilon$ from Eq.(19) arising now {\it only} from 
physics at the scale $M_U$.  Our goal is to attempt to separate out the running of both $\alpha_D$ and $\alpha_{em}$ due to low scale physics from this high scale running of $\epsilon$. 
Given this expression and neglecting the masses of all the light fields as discussed, we can now write the full explicit $s$-dependence of the cross section in the form 
\begin{equation}
\sigma_e(s)=N'~\frac{\alpha_{em}^3 (s)\alpha_D(s)}{s}~{\cal F}(s)\,,
\end{equation}
with $N'$ here being just another number. Measuring $\sigma_e(s)$ at various different energy scales and knowing the energy-dependent behavior of both $\alpha_{em}(s)$ (based on the SM 
and assuming no new charged states below $M_U$) as well as that of $\alpha_D(s)$ (by assumption from the low energy dark sector spectrum), we might, in principle, be able to extract the purely high 
scale dependent ratio shown in Fig.~\ref{fig4}, \ie, the high scale running of $\epsilon$ with the effects of $\alpha_{em}(s)$ and $\alpha_D(s)$ removed, here for the case of time-like $s=Q^2$. 
For completeness, we note that above the electroweak $W/Z//h/t$ mass scales, $\alpha_{em}$ may be always defined in terms of the $SU(2)_L\times U(1)_Y$ running gauge couplings, $g_L,g_Y$, 
via the SM relationship 
\begin{equation}
\alpha_{em}^{-1}=4\pi~\Big(\frac{1}{e^2}=\frac{1}{g_L^2}+\frac{1}{g_Y^2}\Big)\,,
\end{equation}
while $\epsilon$ is, in general,  just a linear combination of the two possible KM parameters above the electroweak scale associated with the DP's mixing with either/both of the (now unbroken) 
$SU(2)_L\times U(1)_Y$ SM's  $W_3$ and $B$ gauge bosons, \ie, $\epsilon=c_w\epsilon_B+s_w\epsilon_W$ \cite{Bauer:2022nwt,Rizzo:2022jti}.

\begin{figure}[htbp]
\centerline{\includegraphics[width=5.0in,angle=0]{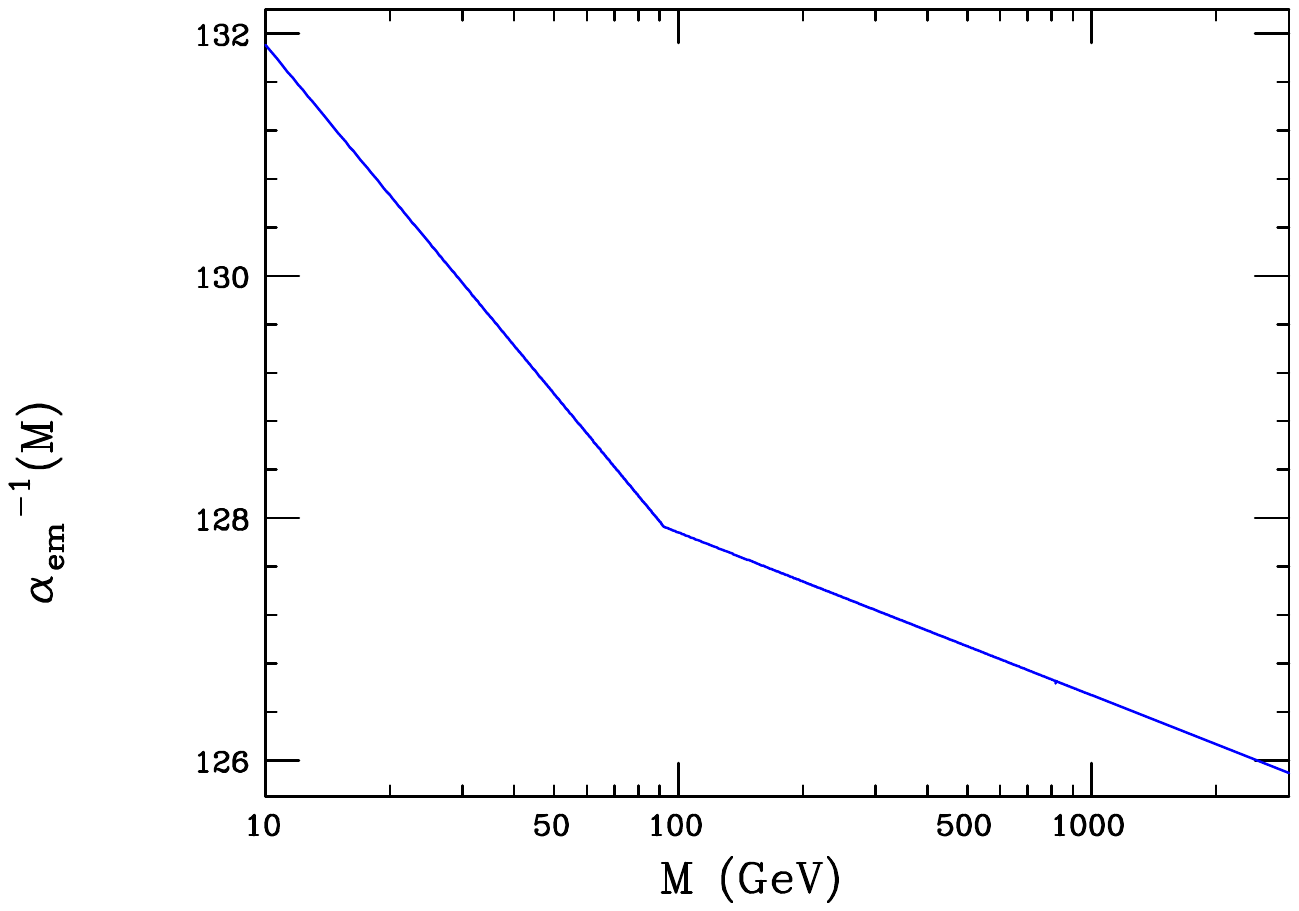}}
\vspace*{-1.3cm}
\caption{Semi-quantitative estimate of the running of $\alpha_{em}$ above the $\sim 10$ GeV $\Upsilon$ mass region and below $M_U$ as described in the text.}
\label{figt}
\end{figure}

Fig.~\ref{fig4} already showed that, quite generally, the effect of the running of $\alpha_D$ can be quite significant. In the case of $\alpha_{em}$, only for purposes of demonstration, we can get a rather 
crude and at best, semi-quantitative 
feeling for its running above $m_\Upsilon \simeq 9.5-10$ GeV (where there are few resonance and thresholds) given that $\alpha_{em}(M_\Upsilon)^{-1}=132.0$\cite{Brambilla:2007cz}  and 
$\alpha_{em}(M_Z)=127.951$\cite{PDG}. We then can employ the lowest order SM RGE's for $g_{L,Y}$ and treating, the $W,Z,h,t$ masses as being very roughly degenerate, obtain the result 
as is shown in Fig.~\ref{figt}. Here we see that the running of $\alpha_{em}$ over the energy range of interest is also seen to be relatively rapid below the $Z$ pole/electroweak scale, but is then 
observed to 
soften significantly above it in the intermediate energy region. However, we note that $\alpha_{em}(s)$ appears to the $3^{rd}$ power in the cross section expression, significantly increasing 
the cross section's sensitivity to this running; thus the combined energy dependence of the overall $\alpha_{em}^3 (s)\alpha_D(s)$ factor in the cross section can be quite significant making the 
high energy sensitivity to the running of $\epsilon$ alone 'difficult' to probe, at best.

Clearly, as we can see from both Figs.~\ref{fig4} and ~\ref{figt}, even if this cross section could be measured, a meaningful extraction of $\epsilon(s)$ would be rather difficult as its 
$s$-dependence is relatively quite weak until the scale $M_U$ is approached rather closely and can easily be overwhelmed by that of both $\alpha_{em}$ and $\alpha_D$. A high-precision 
set of measurements, as well as detailed knowledge of the running of these two gauge couplings would, at the very minimum, be required. Now, of course, the situation is actually {\it much} 
worse than this since the $e^+e^-$ to DM-pair cross section, as is, {\it cannot} be measured since the DM final state is invisible. This type of invisible final state is usually probed via the initial state radiation 
(ISR) of an additional photon, suppressing the cross section and requiring us to understand the SM backgrounds rather well - subject to additional experimental cuts. It is important to be reminded that 
our goal here will not be just to observe the rare signature for this BSM process (which is already difficult as it is $\epsilon^2$ suppressed) but to observe the {\it additional} running of $\epsilon$, apart 
from that due to $\alpha_D$ and $\alpha_{em}$,  as predicted 
by Eq.(19). We will return to these rather important issues below but we first now turn our attention to the analogous, but in many ways much more hopeful, case of the DMom setup.

In the corresponding DMom portal model scenario, employing the same notation as above in the case of scalar (fermionic) DM, the $e^+e^ -\to$ DM pair cross section is instead now given by
\begin{equation}
\sigma_e=\frac{(4)\pi \alpha_D^2}{3s}~\frac{s^2}{(s-m_V^2)^2+\Gamma_V^2m_V^2}~\Big(\frac{s^2}{\Lambda_1^4}\tilde F_1^2+\frac{s}{\Lambda_2^2}F_2^2\Big)\,,  
\end{equation}
where we will again assume we are probing values of $s$ where both $m_V$ and $\Gamma_V$ can be neglected and the PM fields cannot be produced on-shell. 
In the notation of the toy model above, to the order we are working we now simply identify
\begin{equation}
\tilde F_1(s)\simeq f_1(a)+...,~~~~F_2(s) \simeq f_{20}(a)+\frac{s}{12m_E^2}f_{21}(a)+....\,,  
\end{equation}
with $a=m_E^2/m_{W_I}^2$ as before. As noted, it will be frequently useful and simplifying to employ the ratio the next-to-leading to that of the leading contributions to the $F_2$ form factor, 
$r=f_{21}(a)/f_{20}(a)$, in the analysis that follows. For the specific choice of fermionic DM of interest here, this leaves us with
\begin{equation}
\sigma_e=\frac{4\pi}{3}~\alpha_D^2(s)~\Bigg[\frac{s}{\Lambda_1^4}f_1^2+\frac{f_{20}^2}{\Lambda_2^2}\Big(1+r\frac{s}{12m_E^2}\Big)^2\Bigg]= T_1+T_2\,,  
\end{equation}
where the full $s$-dependence is now explicitly displayed once the $\Lambda_i$ are fixed. Recall that within the toy model, since only one value of the mass ratio, $a$, will be realized in actuality, the 
parameters $\Lambda_{1,2}, f_1,f_{20}$ and $r$ are simply $s-$independent numbers to be measured, of which only a subset are fully independent as they are linked together by the toy model 
structure.  

Now one may ask why we do need to keep the NLO term in the $F_2$ expansion; the answer to this can be seen immediately from the factor inside the square bracket in Eq.(28) for the relevant 
cross section. As we will soon see, as a function $\hat s$ (which is simply just $M^2$ in Fig.~\ref{fig6}), apaert from the overall $\alpha_D$ dependence, there is a constant term, proportional 
to $f_{20}^2/\Lambda_2^2$, as well as a linear 
term proportional to the sum $f_1^2/\Lambda_1^4 +rf_{20}^2/(6m_E^2 \Lambda_2^2)$. All terms with the same power of $\hat s$ in the cross section need to be kept for consistency so we are 
forced to include the NLO term proportional to $r$. Now if we use the photon recoil energy as a measure of $M^2=\hat s$, then we can then extract the coefficients of these two terms. As we'll 
see, in the specific toy model realization, however, that the term proportional to $r$ is numerically relatively  small over almost all of the parameter space. If we also assume, as is done here, 
that Belle II can measure the $e^+e^- \to V+\gamma$ cross section, which is also proportional to $f_{20}^2/\Lambda_2^2$, then the three quantities $r, f_{20}^2/\Lambda_2^2$ and 
$f_1^2/\Lambda_1^4$ can be completely determined.

Now as noted, these DM pair production cross sections are by themselves unobservable and the conventional resolution to this problem at $e^+e^-$ colliders is via the emission of an additional ISR 
photon so that the final state appears as a single $\gamma$ balanced by missing energy. To a rather good approximation, in the standard approach, this can be done by turning the $e^+e^-\to$ DM 
cross section into one for a subprocess, multiplying by a photon radiator function and performing the needed phase space integration subject to any experimental cuts. To this end we make use 
of the improved radiator function, ${\cal R}$, as employed in, \eg, Ref.\cite{Chu:2018qrm,Choi:2015zka} and as a first step write the physical $e^+e^-\to \gamma$+DM pair cross section in the form
\begin{equation}
\frac{d\sigma}{dx_\gamma~d\cos \theta_\gamma}=\sigma_e(s\to \hat s)~{\cal R}(x_\gamma,\theta_\gamma)\,,  
\end{equation}
where $\sigma_e$ is given in either Eqs.(19) or (27) depending upon the choice of the KM or DMom setup, with $x_\gamma=2E_\gamma/\sqrt s$ being the scaled energy of the ISR photon so 
that $\hat s=M^2=(1-x_\gamma)s$ is the $e^+e^-\to$ DM pair subprocess center of mass energy squared, and with $\theta_\gamma$ being the photon's scattering angle measured from the $e^-$ 
beam direction in the center of mass frame. The radiator function we use, in the limit of vanishing electron mass, is then just given by\cite{Chu:2018qrm,Choi:2015zka}
\begin{equation}
{\cal R}(x_\gamma,\theta_\gamma)=\frac{\alpha_{em}}{\pi}~\frac{1}{x_\gamma}~\Bigg[\frac{1+(1-x_\gamma)^2}{1-\cos^2 \theta_\gamma}-\frac{x_\gamma^2}{2}\Bigg]\,,  
\end{equation}
so that the integration over $z=\cos \theta_\gamma$ is trivial once the relevant experimental detector angular acceptance cuts, $-z_0\leq z\leq z_0$, are imposed.

\begin{figure}[htbp]
\centerline{\includegraphics[width=5.0in,angle=0]{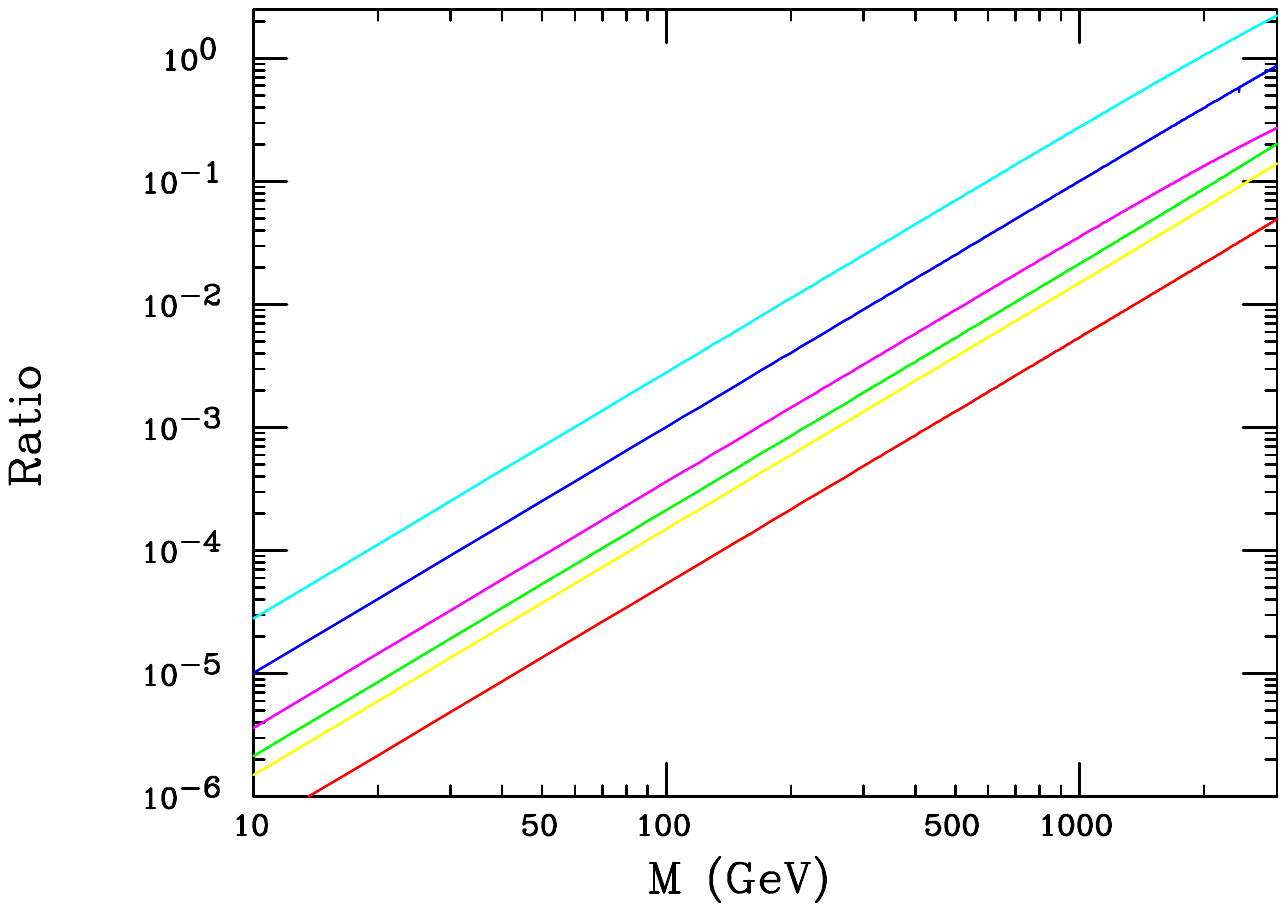}}
\vspace*{-1.3cm}
\caption{The upper three curves are for the quantity $U_2$ with $m_E=1.5,2.5$ and 2.5 TeV assuming also that $a=1/4,1/4$ and 2 from top to bottom, respectively. The lower three curves are for the 
quantity $U_1$ with $m_E=2.5,1.5$ and 2.5 assuming also that $a=2,1/4$ and 1/4, respectively, from top to bottom. Here $M^2=\hat s=s(1-x_\gamma)$ as given in the text.}
\label{fig6}
\end{figure}

Before exploring the implications of the full cross section, it is useful to examine how the various contributions to the subprocess scale relative to each other with differing values of $M=\sqrt {\hat s}$ as 
we alter the requirements on the ISR photon's energy.  In particular, we are interested in the sensitivity to the next to leading terms in the $q^2(=\hat s)/m^2$ expansion relative to the leading 
one employed at low energies since these can be used to probe $M_U$ scale physics. Based on the $2\to 2$ DM pair production subprocess cross section expression above, to this end it is now 
useful to define the two dimensionless quantities
\begin{equation}
U_1=\Big(1+r\frac{\hat s}{12m_E^2}\Big)^2-1,~~~~~~ U_2=T_1(s\to \hat s)/T_2(s\to \hat s)\,,  
\end{equation}
where $U_1$ probes the relative contribution of next to leading term in the dark magnetic dipole form factor compared to the leading term, while $U_2$, with $T_{1,2}$ defined above in 
Eq.(27), probes the relative contribution of $\tilde F_1$ and $F_2$ terms to the DM pair production subprocess cross section. Larger values of $U_{1,2}$ would clearly indicate places in the toy model 
parameter space where the 
next to leading order terms in the expansion become important. Note also that both $U_{1,2}$ are independent of the value of $\alpha_D$ as this quantity appears 
only as an overall factor in the subprocess cross section. Fig.~\ref{fig6} shows the values of both $U_{1,2}$ as function of $M=\sqrt {\hat s}$ for different chosen values of the parameter 
$a=m_E^2/m_{W_I}^2$ and of $m_E$, not far above those that might be excluded by null searches at the HL-LHC as discussed above. As might be expected, the largest values of $U_{1,2}$ 
appear when $E_\gamma$ is small so that $M$ is as large as possible, \ie, when more momentum flows into the subprocess and not into the emitted photon. However, even for values of 
$M$ approaching $\sim 1$ TeV, which will clearly still remain inaccessible to the first generation of any planned $e^+e^-$ colliders, these quantities, unfortunately, never get very large and are always less 
than, \eg, $U_2\sim 0.1-0.2$ and with $U_1$ sometimes as small as $\sim 0.005$. Thus, although there is some apparent sensitivity to the next to leading terms it may not be very significant depending 
upon our chance location in the toy model - or any realistic model's - parameter space and higher collision energies are clearly very helpful in this regard. However, we also see that, at least in this 
toy model, since $U_2$ is always somewhat larger than $U_1$, it is the leading term in $\tilde F_1$ which is seen to dominate over the sub-leading term in $F_2$ in terms of sensitivity to next to leading 
order effects. 

We stress again that it is obviously advantageous - certainly in the DMom setup - to go to very large values of $\sqrt s$: not only are we much closer to the PM scale that we wish to probe 
but we also observe that while any of the SM backgrounds will like fall semi-quantitatively as $\sim 1/s$, the signal does {\it not} until the PM scale is reached and/or surpassed.  This will be explored 
further below.

\section{Radiative Dark Matter Signals at Future $e^+e^-$ Colliders?}

As a first step in extracting information about $M_U$-scale physics, we must address a much more basic issue: whether or not we will be able to see any BSM signal at all, not to mention the 
particular next to leading order effects we're after.  If so, the important next step is to ask if the next-to-leading contributions can be disentangled from those of the leading term. If not, then we really 
do not gain any new information about high scale physics. 
Although we already have learned that this is a difficult problem at best, now we must examine searching for these non-leading effects at $e^+e^-$ colliders a bit more realistically. One obvious issue is 
the much-studied SM background to the $e^+e^-\to \gamma +{\rm nothing}$ process that we're examining which arises from both $s$-channel $e^+e^-\to Z\to \bar \sum_i \nu_i \nu_i$ as well as via the 
$t-$channel, $W$-exchange process,  $e^+e^-\to \bar \nu_e \nu_e$, both with an additional photon emission. Useful analytical expressions for these SM backgrounds can be found, \eg, in 
Refs.\cite{Hirsch:2002uv,Barranco:2007ej,Berezhiani:2001rs,Escrihuela:2019mot} that we will employ in the analysis below. Fortuitously (or not), the $W$ process is dominant but it is also purely 
left-handed and so can be at least be partially suppressed by employing a suitable choice of $e^\pm$ beam polarizations. 
Also, one finds that the same polarization choice that reduces this $W$-induced SM background also slightly enhances the ratio of the $\tilde F_1$ to the $F_2$ contribution to the DM 
cross section. This is easily seen as a spin-1, pure vector $s$-channel exchange interaction, such as that proportional to $\tilde F_1$, couples equally to the helicities $e_L^-e_R^+$ and 
$e_R^-e_L^+$, while the $s$-channel, magnetic dipole interaction proportional to $F_2$ couples equally to the helicity combinations $e_L^-e_L^+$ and $e_R^-e_R^+$ due to the additional 
$\gamma$-matrix in the coupling structure. Assuming beam polarizations of $|P_{e^-}|=0.8$ and $|P_{e^+}|=0.3$, as may be possible at the linear colliders such as ILC, CLIC or 
C$^3$\cite{ILC:2007oiw,ILCInternationalDevelopmentTeam:2022izu,Aicheler:2012bya,Bai:2021rdg}, one can enhance the ratio of the $\tilde F_1$ to 
the $F_2$ contributions to the cross section by a factor of $(1+0.8\cdot 0.3)/(1-0.8\cdot 0.3)=1.63$. Simultaneously, a purely LH-coupling, \eg, $e_L^-e_R^+$, as is that induced by the 
SM $W$ is then suppressed by a factor of $(1-0.8)(1-0.3)=0.14$ while, correspondingly a purely RH-coupling, \eg, $e_R^-e_L^+$, which occurs as part of the much smaller SM $s$-channel $Z$ 
exchange, would be enhanced by a factor of $(1+0.8)(1+0.3)=2.34$. 

\begin{figure}[htbp]
\centerline{\includegraphics[width=5.0in,angle=0]{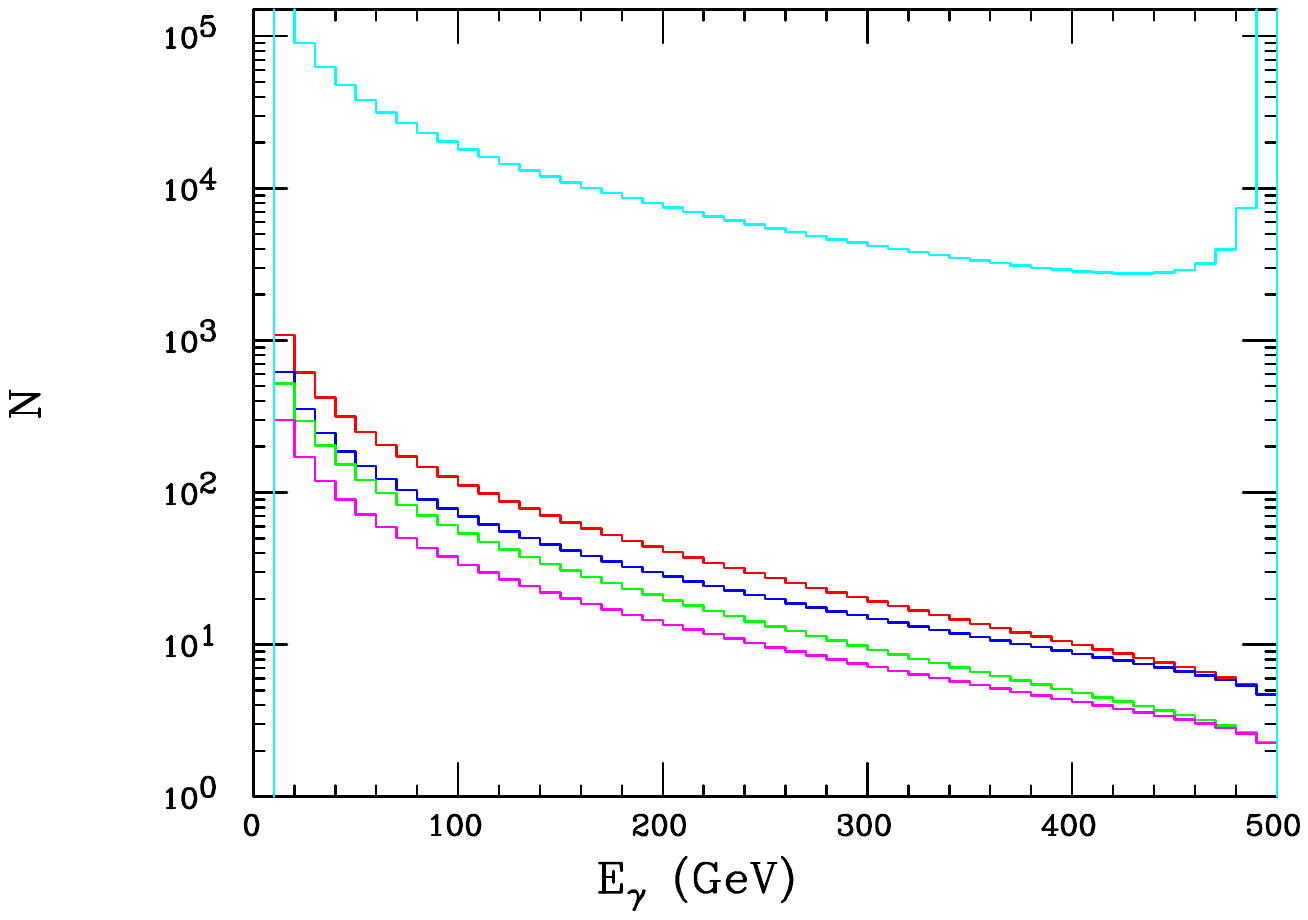}}
\vspace*{-1.3cm}
\caption{Signal and SM background (cyan) event rates in 10 GeV photon energy bins for the process $e^+e^-\to \gamma +$nothing arising from fermionic DM production assuming $\sqrt s=1$ TeV, an 
integrated luminosity, $L$, of 1 ab$^{-1}$ with $P_{e^-}=0.8,P_{e^+}=0.3$, $z_0=0.995$, $E_\gamma>10$ GeV and the DMom model parameter choices $a=1/4$, $m_E=1.5$ TeV, to demonstrate the 
effect for which we are searching for purposes of demonstration. The red and blue (green and magenta) histograms assume values of $\alpha_D({\rm 1~ TeV})=0.30(0.25)$ which run to lower 
energies as $E_\gamma$ increases, while the the blue and green histograms assume that only $f_{20}\neq 0$.}
\label{fig7}
\end{figure}

Although all of these polarization factors do work in our favor, the problems we face are still quite formidable. The extraction of new physics signals in monophoton events at lepton colliders is a well-studied 
problem in the literature (see, \eg, Refs.\cite{Chae:2012bq,Kalinowski:2021tyr,Khalil:2021afa,Black:2022qlg,Dreiner:2012xm,Bauer:2023loq}).  Consider, as an example, the results shown 
in Fig.~\ref{fig7} where we assume that $\sqrt s=1$ TeV (as the relevant contributions grow with $\sqrt s$) for the pseudo-Dirac DM scenario with the toy model parameters chosen to show a 
potentially `significant' 
effect, here for an integrated luminosity, $L$, of 1 ab$^{-1}$. We will further require a photon energy cut $E_\gamma \geq 10$ GeV to act as a trigger and we also assume a 
detector coverage down to small angles, \ie, $|z|\leq z_0=0.995$. Employing the expressions in Refs.\cite{Hirsch:2002uv,Barranco:2007ej,Berezhiani:2001rs,Escrihuela:2019mot} for the SM 
background as well as making the beam polarization assumptions above to minimize the SM background (and to increase the signal rate) leads to the cyan histogram for the still quite large 
background found in this Figure. The significantly smaller DM signal in this setup in the seen in lower set of four histograms. Note that in performing these calculations we have assumed that 
$\alpha_D({\sqrt {\hat s}=\rm 1~ TeV})$ takes on value of 0.30(0.25). While these $\alpha_D$ values for the signal may be close to applicable on the left-hand side of the plot, as $E_\gamma$ increases, 
${\sqrt {\hat s}}$ decreases so that the $\alpha_D(\sqrt {\hat s})^2$ pre-factor in the cross section {\it also} decreases as a result of the energy dependence discussed above. Due to this effect, for the 
chosen pseudo-Dirac DM case shown here, these cross sections are gradually reduced by an additional factor of $\sim 2$ as a result of running once large values of $E_\gamma$ are reached.  
Of course the {\it ratios} of the next to leading to leading order production rates are completely unaffected by this running of $\alpha_D$ since it is an overall cross section pre-factor. Now recall that 
our goal is {\it not} simply to observe the signal but to {\it differentiate} the full next to leading order result, where $f_{20},f_{21},f_1$ are
all non-zero, from the leading order case where only $f_{20}$ is finite. Examining these histograms, we see that this means {\it measuring} the signal with rather high precision, which  
seems - just at the level of statistics - to generally be rather dubious given the huge SM background, even with the optimistic assumptions we've made above and especially when the considerations of the 
numerous possible systematic errors (luminosity, polarization, \etc.) have not been included in our estimates.

Naturally, we might imagine that going to a different, more judiciously chosen point in the toy model parameter space, \eg, $a=1$ for the same value of $m_E=1.5$ TeV, would make things `easier'; 
as we can see for Fig.~\ref{fig5}, this should indeed be the case as $f_1, f_{20}$ and $r$ would all increase as the value of $a$ is also increased. Fig.~\ref{fig7pp} shows the result of this parameter shift 
where the DM cross sections are all increased by, very roughly, a factor of $\simeq$ 8-10; here we see that we might have a reasonable chance to observe the contributions of the next to leading 
order terms. Clearly, there will exist at least {\it some} regions of the DMom toy model parameter space wherein we will be able to see such terms. Of course, one will need to make a more sophisticated 
set of calculations to which we note that other potentially important effects (such as beamsstrahlung, higher order radiative corrections as well as the errors induced by the other systematics effects 
mentioned above that have been ignored here) would be needed to be included. Including detector level effects would also be important to reach a completely firm conclusion.

\begin{figure}[htbp]
\centerline{\includegraphics[width=5.0in,angle=0]{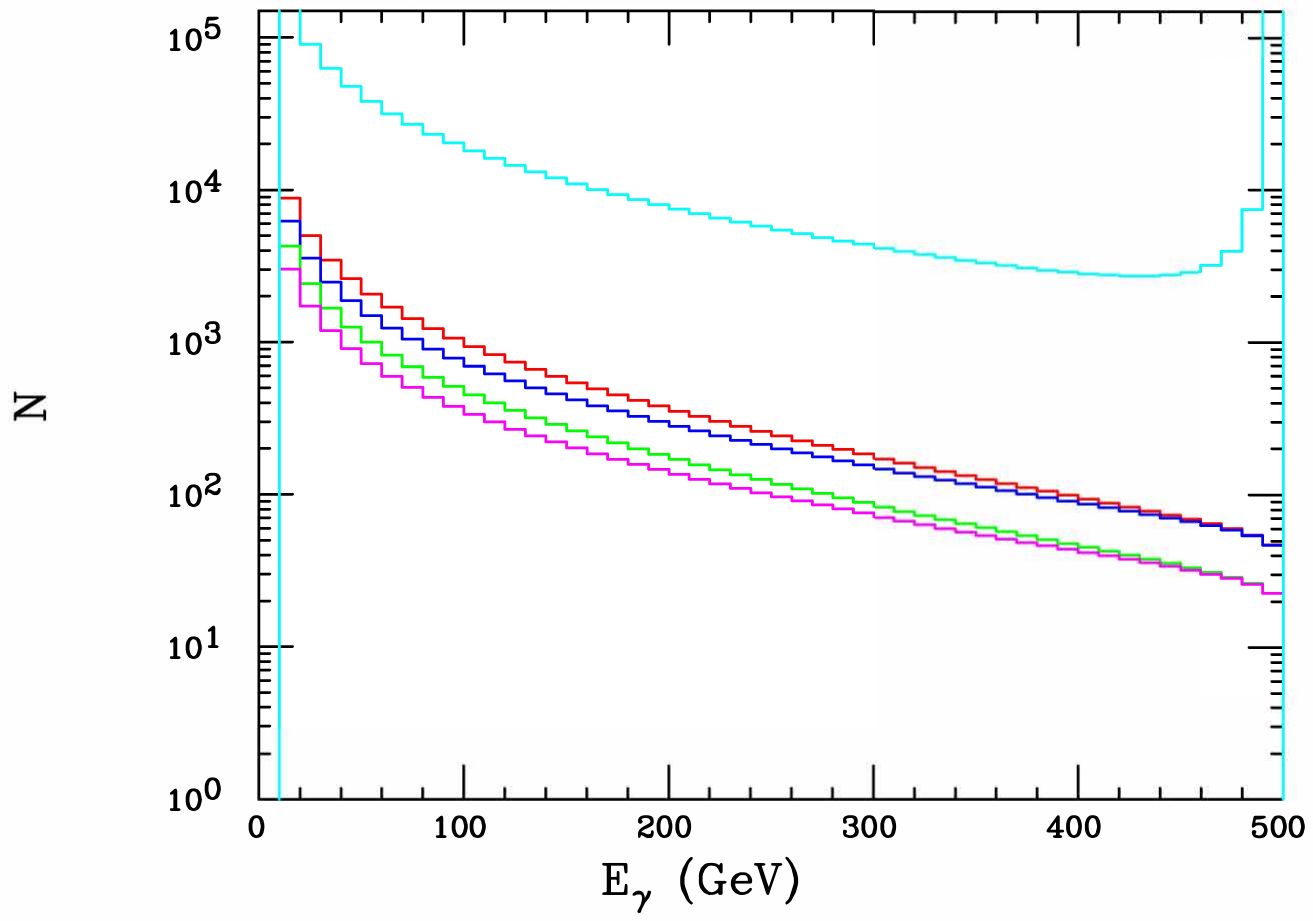}}
\vspace*{-1.3cm}
\caption{Same as the previous Figure but now assuming instead that $a=1$.}
\label{fig7pp}
\end{figure}

To do significantly better {\it without} changing the chose values of the toy model parameters, \eg, $a=1/4 \to a=1$, one could also move to larger values of $\sqrt s$, \eg, 2 TeV, \ie, closer to the PM 
mass threshold, as then the signal over background ratio would increase and the values of both $U_{1,2}$ would also increase due to their stronger $s$-dependence.  Of course, at such a parameter 
point we're almost at the scale $M_U$, which for the cases considered here is essentially at 3 TeV or a bit 
larger.  Fig.~\ref{fig7p} shows the result of the move to such higher energies accompanied by an simultaneous increase in the integrated luminosity, but still ignoring other effects here such as 
beamsstrahlung. Of course we must begin such considerations with a warning:  due to the proximity of some of the $\sqrt {\hat s}$ values that arise in this calculation to the $2m_E$ threshold at 3 TeV, 
we may wonder whether, in this case, the truncation of our form factor expansion to (just) the next to leading order terms is actually warranted and that even higher order terms must be included at 
the level of precision in which we are interested. Be that as it may, we can still ask whether of not these NLO terms are observable relative to just the LO ones; based on the results in Fig.~\ref{fig7p}, 
it would seem that this may be possible provided that the effects we've neglected are not of extreme importance. However, again 
a more realistic simulation study is required to confirm (or deny) these tentative 
conclusions.

\begin{figure}[htbp]
\centerline{\includegraphics[width=5.0in,angle=0]{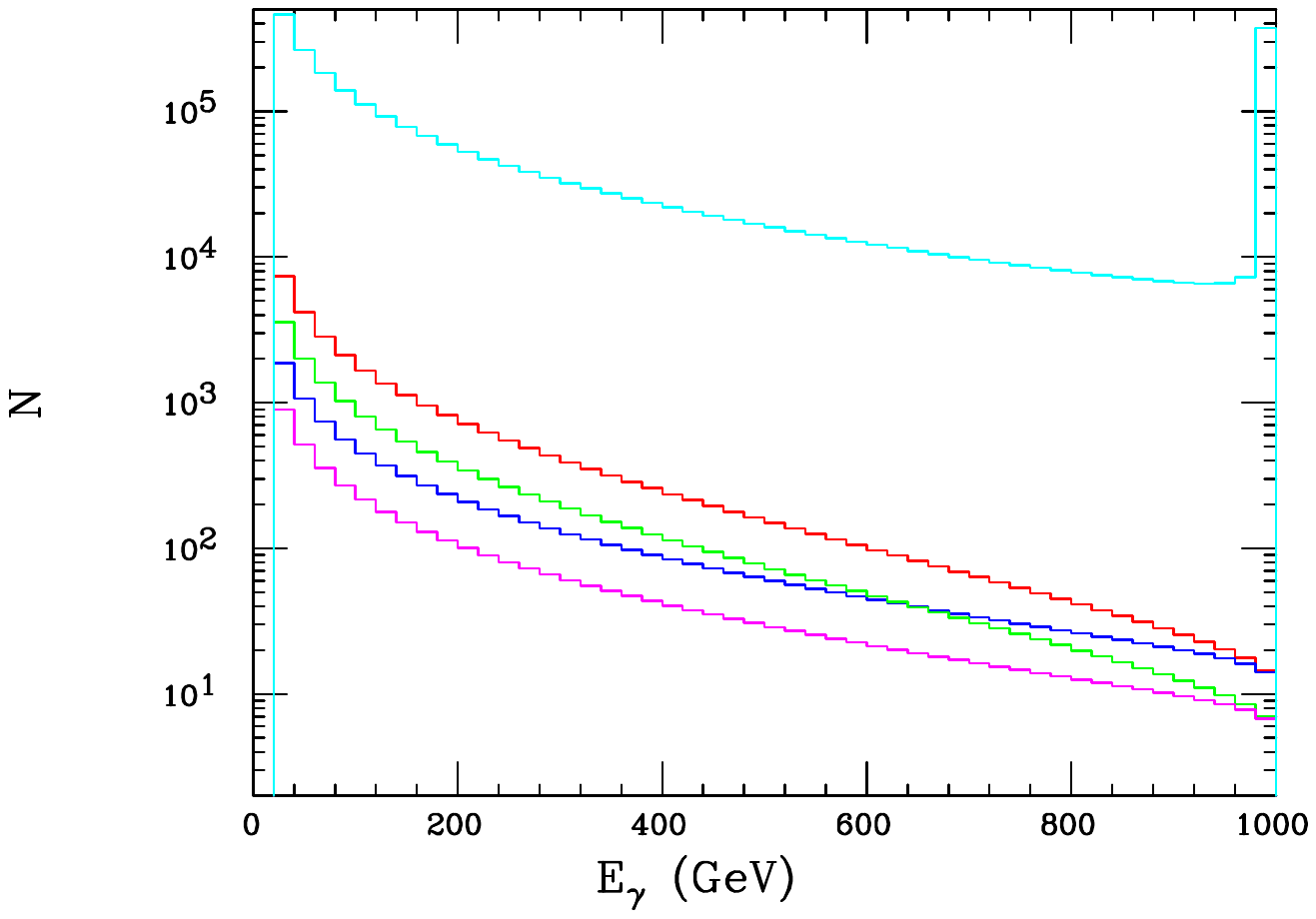}}
\vspace*{-1.3cm}
\caption{Same as the previous Figures but now assuming that $\sqrt s=2$ TeV and 20 GeV bins of $E_\gamma$ with $a=1/4$ and an integrated luminosity of $L=3$ ab$^{-1}$. Here we also require 
that $E_\gamma >20$ GeV to act as a trigger.}
\label{fig7p}
\end{figure}

Unlike the DMom setup wherein the $e^+e^-\to $DM cross section essentially grows with energy (relative to the usual $1/s$ SM behavior) until the PM fields and associated gauge 
bosons are resolved, in the KM case the corresponding cross section {\it always} remains crushed by the $\epsilon^2$ prefactor. We would then anticipate it to {\it yield} very highly suppressed 
signatures at $\sqrt s=1$ TeV when the DP is not produced on-shell, which is the case of interest. Thus, we should expect the somewhat hopeful showing found for the DMom setup for certain parameter 
choices to be made to appear very promising by comparison to that in the KM model. Recall, we are not simply looking for a signal of new physics but the ability to differentiate the case of explicit running 
for $\epsilon$, wherein the PM scale sensitivity lies, to that where the running of $\alpha_{em}$ and $\alpha_D$ control this entirely, \ie, ${\cal {F}}\neq 1$, in the notation above. The results displayed 
in Fig.~\ref{fig8} show that it is almost certain that this process in the KM setup will never be observed in the case when $V$ is far off-shell at a high energy $e^+e^-$ collider, being always 
buried under the SM background by over six orders of magnitude.

\begin{figure}[htbp]
\centerline{\includegraphics[width=5.0in,angle=0]{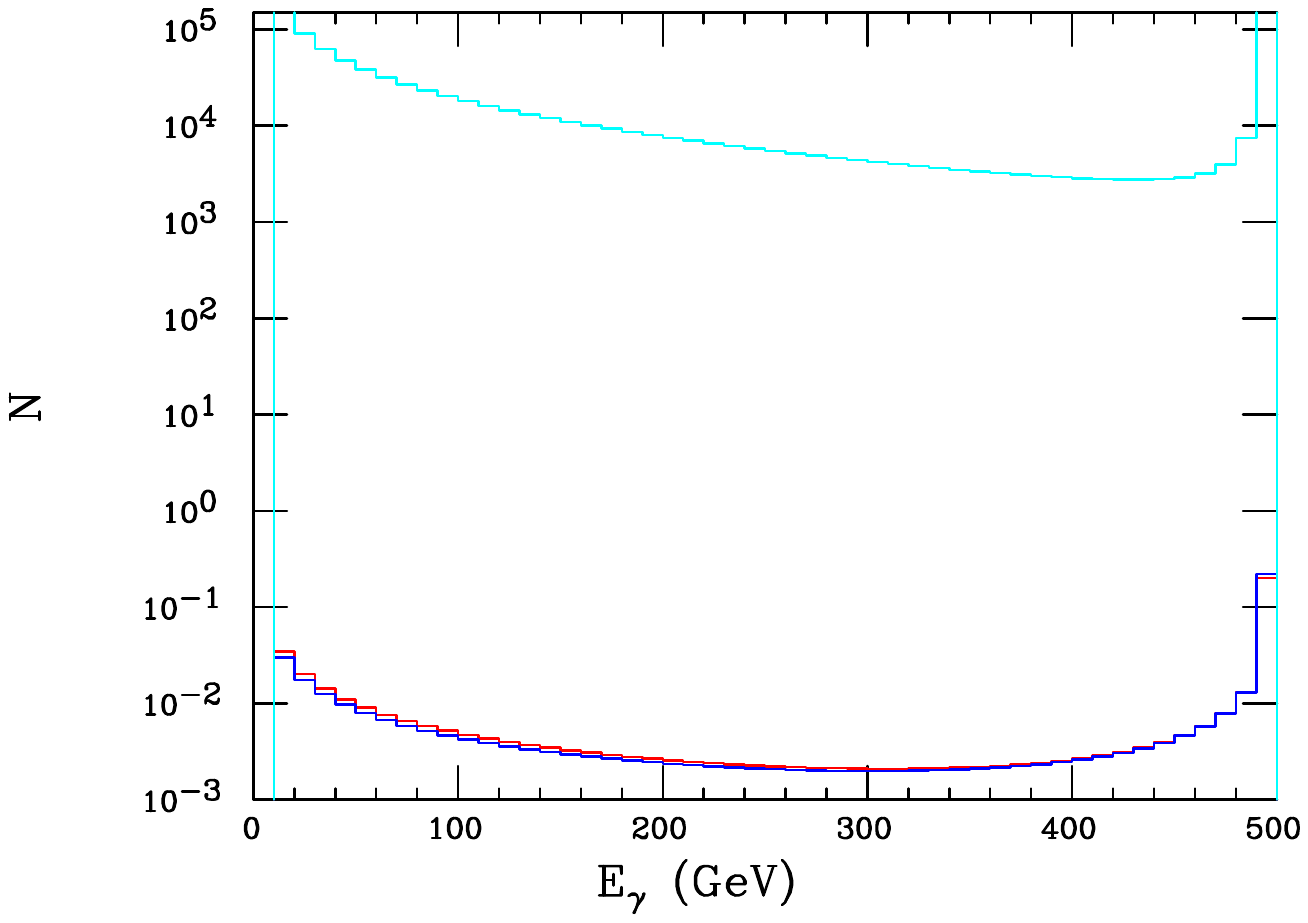}}
\vspace*{-1.3cm}
\caption{Similar to Fig. 8, showing the same SM background (in cyan) under identical assumptions. However, the two lower histograms display the result for the KM model as 
discussed in the text both including (red) or ignoring (blue) the explicit running of $\epsilon$. Here $\epsilon_0=3\cdot 10^{-4}$ has also been assumed; other assumptions and parameters 
choices are as in the previous Figures.}
\label{fig8}
\end{figure}

\section{Standard Model Signals at $e^+e^-$ Colliders?}

If electrons are not the only SM charged fermions that have dark multipole moments, we can imagine, \eg, processes such as $e^+e^-\to \bar ff$ or $\bar qq\to e^+e^-$ mediated via a DP and with 
DMom vertices appearing at both ends, analogous to a double-peguin diagram{\footnote {Of course, in a more serious and complete theory one can imagine several other additional diagrams 
representing related processes which may also make potentially important 
contributions.}}. In the corresponding situation in the KM model, such, now tree-level, processes will also exist with DP exchange but are very highly suppressed by a factor of $\epsilon^4\sim 10^{-14}$ 
so would, of course, be completely unobservable. In the DMom setup, the relative growth in the effective couplings with $\sqrt s$ below the scale $M_U$ would instead potentially compensate for the 
tiny coupling that is required at very low energies to obtain the observed value of the DM relic density. 
However, we recall that the $F_2$ magnetic dipole tensor structure, in the limit of massless fermions, does {\it not} interfere with the usual vector and axial-vector couplings appearing in  the 
$\gamma,Z$ exchanges present in the 
SM so can only contribute quadratically in the expression for the resulting cross section. This means that such a term in the squared amplitude corresponding to DP exchange (in the $m_V\to 0$ limit) 
would appear proportional to the quantity (in obvious notation for $e^+e^-\to \bar ff$)
\begin{equation}
d_2=\frac{\alpha_D^2}{\alpha_{em}^2}~\Bigg(\frac{ sF_{2e}^2}{\Lambda_{2e}^2}\Bigg)~\Bigg(\frac{sF_{2f}^2}{\Lambda_{2f}^2}\Bigg)\,,  
\end{equation}
which takes the form of a dim-8 contact term contribution. On the other hand, the $\tilde F_1$ term, when coupling to conserved currents (as is the case for the massless fermion assumed here), 
appears similar to an energy-dependent vector coupling, $v_f(V)\sim g_D s \tilde F_{1f}/\Lambda_{1f}^2$, for the fermion, $f$, to the essentially massless DP so that it would interfere with the 
SM contributions in the squared amplitude, appearing only linearly but also necessarily for {\it both} $e$ and $f$ simultaneously.  In the same case of $s$-channel DP exchange, the products of 
these two coupling structures, \eg, 
\begin{equation}
d_1=\frac{\alpha_D}{\alpha_{em}}~\Bigg(\frac{s\tilde F_{1e}}{\Lambda_{1e}^2}\Bigg)~\Bigg(\frac{s \tilde F_{1f}}{\Lambda_{1f}^2}\Bigg)\,,  
\end{equation}
always appears in this interference term, scaled by squares of the corresponding SM $\gamma$ and $Z$ electroweak couplings, \etc, and we are again led to another effective dim-8 contact term 
contribution to the 
squared amplitude.  Thus we conclude that the leading DMom contribution to `conventional' $e^+e^-$ annihilation cross sections would necessarily appear first at dim-8 and so the search sensitivity 
for these new interactions would likely be somewhat restricted in comparison to the more commonly explored dim-6 contact terms. As is well-known\cite{Rizzo:2003sz}, for a statistics limited 
search, the mass reach for a dim-8 operator scales roughly as $\sim (s^3L)^{1/8}$ whereas for a dim-6 operator one instead finds the scaling $\sim (sL)^{1/4}$.

As in the previous Section, we must first address a primary issue: are the contributions of these effective operators potentially visible in the experimental data? 
We note for clarity and consistency that we limit ourselves to just the contributions from dim-8 operators (and not higher) so that we will consider only those terms arising from $f_{20}$ and $f_1$. As 
$f_{20}$ appears as the chief contributor to the low energy observables, it is the additional contribution due to $f_1$ to these SM $e^+e^-$  processes that we seek and wish to extract in the 
current discussion since only it can provide additional information about $M_U$-scale physics. Of course, the actual numerical effect of $f_{21}$, if it were to be included, would be at most of order a 
$\sim 2\%$ shift in the apparent value of $f_{20}$ so is, in practice, ignorable here. The main issues we need to address then are whether or not the dim-8 effects are visible at all and if so are those 
arising from $\tilde F_1$ separately visible, over and above, those arising from $F_2$. Affirmative answers to these questions are necessary before any high scale physics information can be 
usefully extracted.

Attempting to evaluate the expressions above for $d_{1,2}$ in our toy model for an arbitrary SM fermion, $f$, would seem problematic as then we would need to know if $f$ does or does not have a 
PM partner which determines how it would interact with the $W_I$, if at all. Exactly the same problem would arise for the analogous study of, \eg,  the $\bar q q$ initiated Drell-Yan process at 
the LHC or FCC-hh.  Of course, in a more realistic model framework, the answer to this question is clearly known from the model structure but not so in the toy model framework employed here. 
However, to be a bit conservative, if one assumed that the toy model respected $e-\mu$ universality, then we partially know the answer to this question and also that the number of relevant free 
parameters would, at least, be no greater than in our consideration of the DM production process in the previous Section. Additionally, one could avoid this issue altogether and just consider 
Bhabha scattering as then the electron is the only SM particle involved. We now examine both of these two possibilities in turn.

Paralleling, \eg, Refs.\cite{Pasztor:2001hc,Pankov:2004ma}, we can in this case write the $e^+e^-\to \mu^+\mu^-$ polarized cross section in the DMom setup in obvious notation as just
\begin{equation}
\frac{d\sigma}{dz}=\frac{\pi \alpha_{em}^2}{2s}~\Bigg(\Big[\frac{u^2}{s^2}A_++\frac{t^2}{s^2}A_-\Big](1-P^-P^+)+\frac{u^2}{s^2}(P^+ -P^-)A_P+(1+P^-P^+)~d_2\frac{(t-u)^2}{s^2}\Bigg)\,,  
\end{equation}
where, as usual in the massless limit, $t,u=-s(1\mp z)/2$, and where 
\begin{equation}
A_+=|A_{LL}|^2+|A_{RR}|^2, ~~~~A_-=|A_{LR}|^2+|A_{RL}|^2,~~~~A_P=|A_{LL}|^2-|A_{RR}|^2\,,  
\end{equation}
and the individual amplitudes are given by
\begin{equation}
A_{LL}=1+pg_L^2+d_1, ~~~~A_{RR}=1+pg_R^2+d_1,~~~~A_{LR}=A_{RL}=1+pg_Lg_R+d_1\,,
 \end{equation} 
with $d_{1,2}$ given above by taking $f\to e$ and where we have defined the familiar SM quantities
\begin{equation}
g_{[L,R]}=\frac{[-1/2+x_w,~x_w]}{\sqrt {(x_w(1-x_w)}}, ~~~~~p=\frac{s}{s-m_Z^2}\,,  
\end{equation}
which is adequate as we're far away from the $Z$-pole. Note that since $d_1$ reflects a (higher-dimension) vector-like coupling it contributes equally to all the amplitudes. Following the same arguments as 
in the last Section, since $d_2$ is `known'  from low energy measurements and the DM relic density, our goal here is to establish the effects of $d_1$ over and above those arising from just $d_2$ alone 
as this will provide additional information on $M_U$-scale physics. 

\begin{figure}[htbp]
\centerline{\includegraphics[width=5.0in,angle=0]{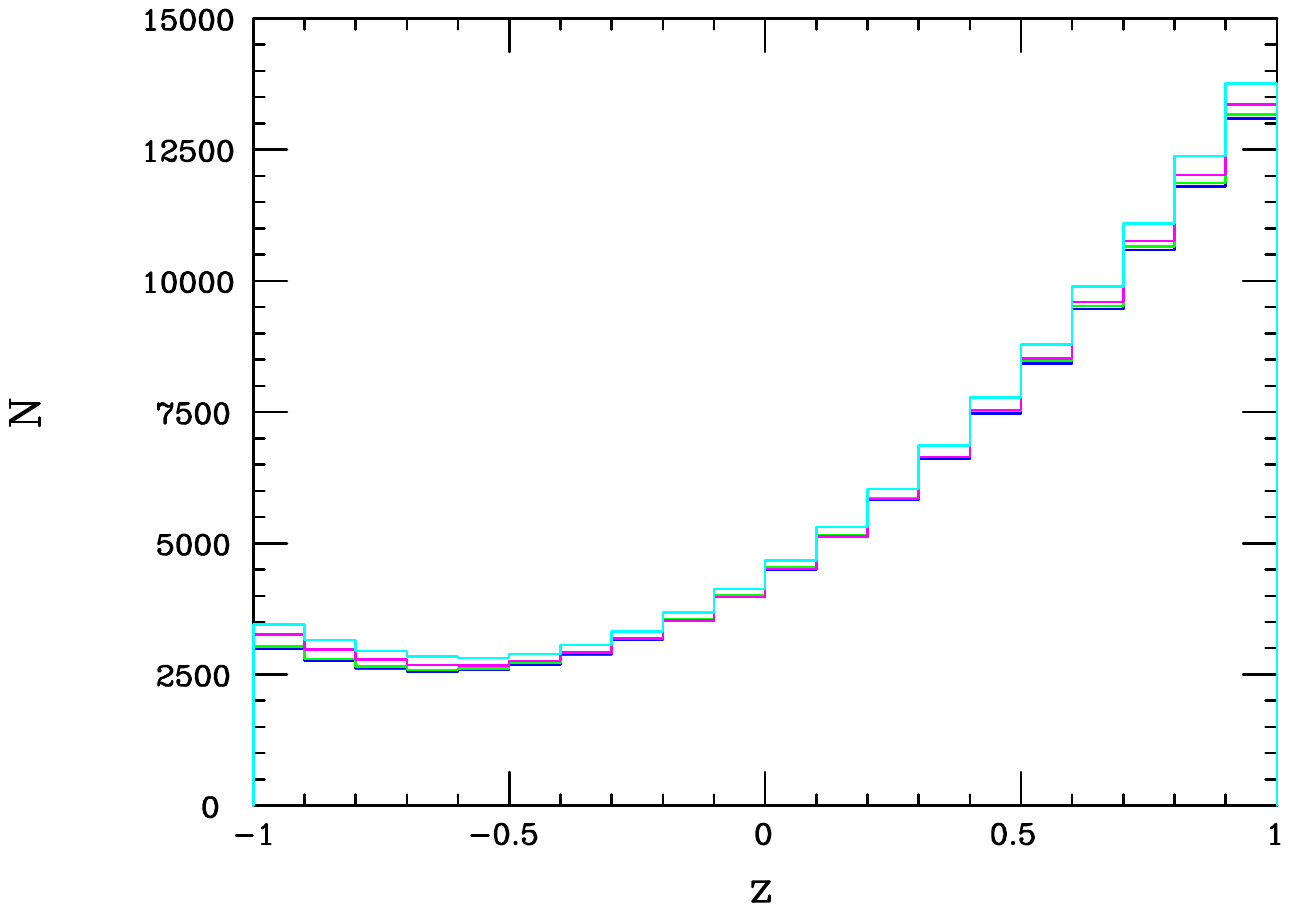}}
\vspace*{-0.8cm}
\centerline{\includegraphics[width=5.0in,angle=0]{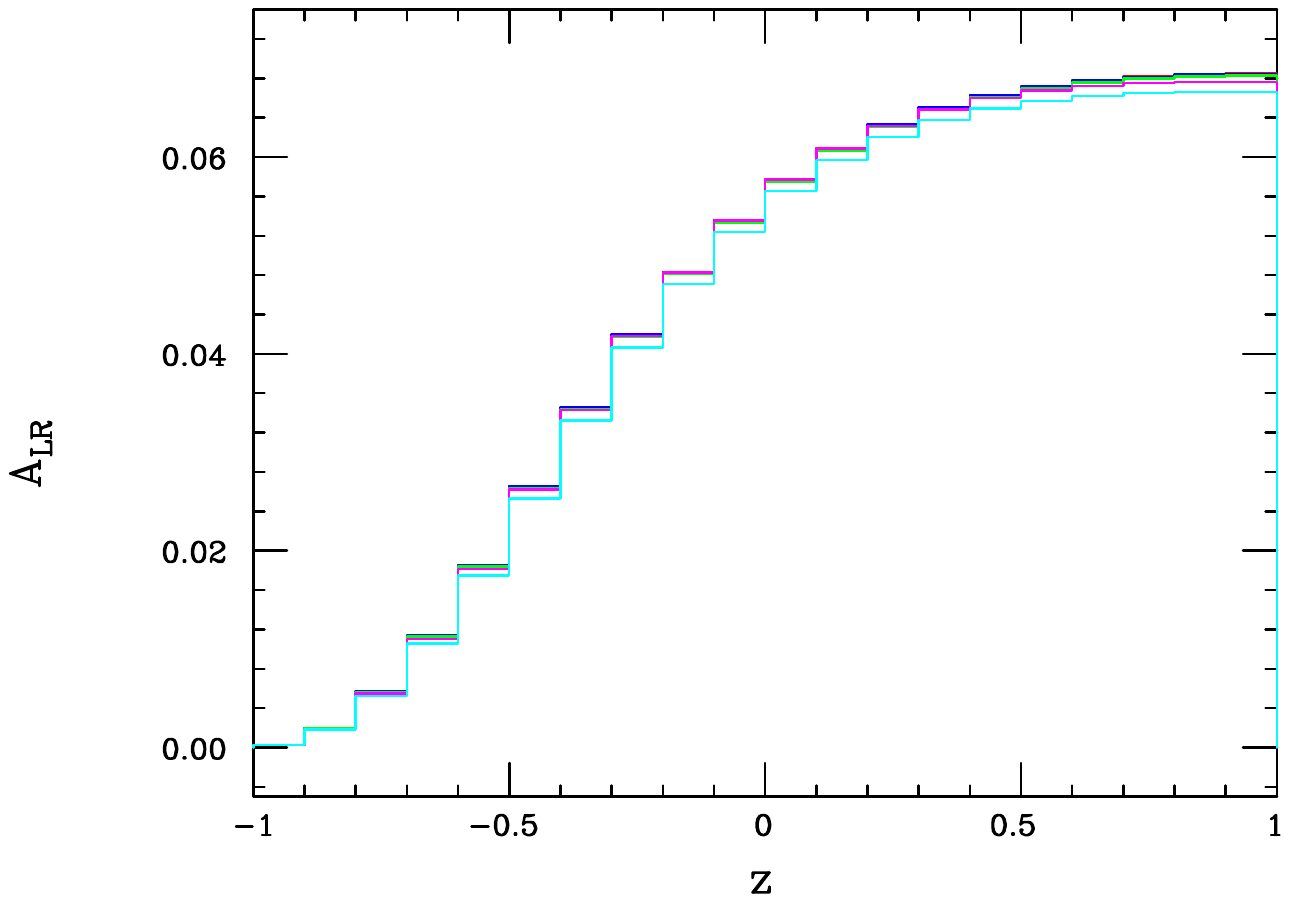}}
\vspace*{-1.3cm}
\caption{(Top)  The event rate for the process $e^+e^-\to \mu^+\mu^-$ at $\sqrt s=1$ TeV assuming unpolarized beams and an integrated luminosity of 1 ab$^{-1}$. The red histogram is the SM 
prediction while the green (blue) histograms are for the toy model benchmark $a=1/4$, $m_E=1.5$ TeV, neglecting (including) the contribution from $\tilde F_1$, respectively. The magenta 
(cyan) histograms are for the corresponding $a=1$ benchmark again neglecting (including) the contribution from $\tilde F_1$, respectively. (Bottom) The left-right polarization asymmetry, $A(z)$, 
as described in the text, for the same cases and labels as in the Top panel. For the DMom predictions, $\alpha_D(1~ {\rm TeV})=0.3$ has been assumed.}
\label{figmu}
\end{figure}

The top panel in Fig.~\ref{figmu} shows the result for this cross section in the case of unpolarized beams in the SM at $\sqrt s=1$ TeV with an integrated luminosity of 1 ab$^{-1}$, contrasted with those 
for the two toy model DMom benchmark cases previously considered, \ie, $a=1/4,1$ with $m_E=1.5$ TeV, both with and without the next to leading order terms from $\tilde F_1$ contained in $d_1$   
included in the calculation. As in the last Section, effects such as beamsstrahlung - which will somewhat degrade the results displayed here - have been ignored in these presentations but can they 
be included at an approximate level as part of the analysis together with such systematic effects as the luminosity uncertainty. We find that the relative shapes of these distributions are a more powerful 
discriminator than is the overall muon-pair production rate which mainly suffers from the previously mentioned respectable luminosity uncertainty which here is assumed to be $0.25\%$, although a 
reduction in this value by roughly a factor of $\simeq 2$ may likely be possible at the ILC\cite{ILCInternationalDevelopmentTeam:2022izu}. Staring at this plot, we see that the two $a=1/4$ cases are 
visually indistinguishable from the SM, or each other, being only a fraction of $1\sigma$ apart even in the absence of beamsstrahlung. The two $a=1$ cases, on the other hand, appears by eye to be 
more hopeful and we find that the $d_1=0$ and $d_1$ non-zero cases are indeed distinguishable at the $\simeq 5\sigma$ level even when beamsstrahlung effects are (approximately) accounted 
for, although a more realistic detector-level analysis, including radiative corrections to the tree-level processes here, is certainly warranted to confirm this result. Obviously also, in a realistic model, 
the other possible contributions that we've omitted here would also need to be included.

Using the above cross section expression, since the $e^\pm$ beams are polarized,  we can also form the usual angular-dependent left-right polarization asymmetry observable{\footnote {For a recent survey of $e^+e^-$ observables for indirect new physics signatures, see Ref.\cite{Das:2021esm}.}}, defined as usual by 
\begin{equation}
A(z)=\frac{d\sigma(P^+,P^-)-d\sigma(-P^+,-P^-)}{d\sigma(P^+,P^-)+d\sigma(-P^+,-P^-)}\,,  
\end{equation}
here taking the values $P^-=-0.8$ and $P^+=0.3$ as employed in the previous Section.  From the bottom panel in Fig.~\ref{figmu}, we see that the predictions are all visually extremely close. Allowing for 
a 0.003 polarization uncertainty\cite{ILCInternationalDevelopmentTeam:2022izu}, we find significantly less than a $1\sigma$ distinguishability arising solely from this quantity. This is not overly 
surprising as the $d_1$ 
coupling is itself vector-like; in addition, dim-8 effects were difficult to observe employing this observable for the $\mu$-pair production process in other previously examined dim-8 physics scenarios, \eg, in 
Refs.\cite{Hewett:1998sn,Rizzo:1998fm} for ADD gravity\cite{Arkani-Hamed:1998jmv}.

We can repeat our cross section study for the case of Bhabha scattering which `suffers' from a very large SM (\ie, photon) pole in the forward direction as can be seen in Fig.~\ref{fig10} where the very 
forward region has been judiciously removed to magnify the backward region. This is both a blessing and a curse as it leads to much higher event rates in that direction {\it but} also with a reduced 
sensitivity to any new physics due to the dominance of this same photon pole. To a reasonable extent, here the added statistics wins, even if one cuts away the very forward region and accounts for 
both luminosity and (at least approximately) beamsstrahlung effects as we had done above for the case of the $\mu^+\mu^-$ final state. In the present case, one finds that the $f_1=0$ and 
$f_1 \neq 0$ results are easily distinguishable at $\sim 5\sigma$ or more (though not by eye) 
for {\it both} the $a=1/4$ and $a=1$ parameter choices. Of course, as in the case of $\mu$-pair production, a detailed detector-level study with higher order radiative effects included needs to be 
performed to validate (or not) these tentative conclusions. However, this result may not be too surprising overall as in the dim-8 ADD study of Bhabha scattering\cite{Hewett:1998sn,Rizzo:1998fm}, 
the Bhabha process led to comparable model constraints as did all the other $\bar ff$ final states combined due to the enlarged statistics.

\begin{figure}[htbp]
\centerline{\includegraphics[width=5.0in,angle=0]{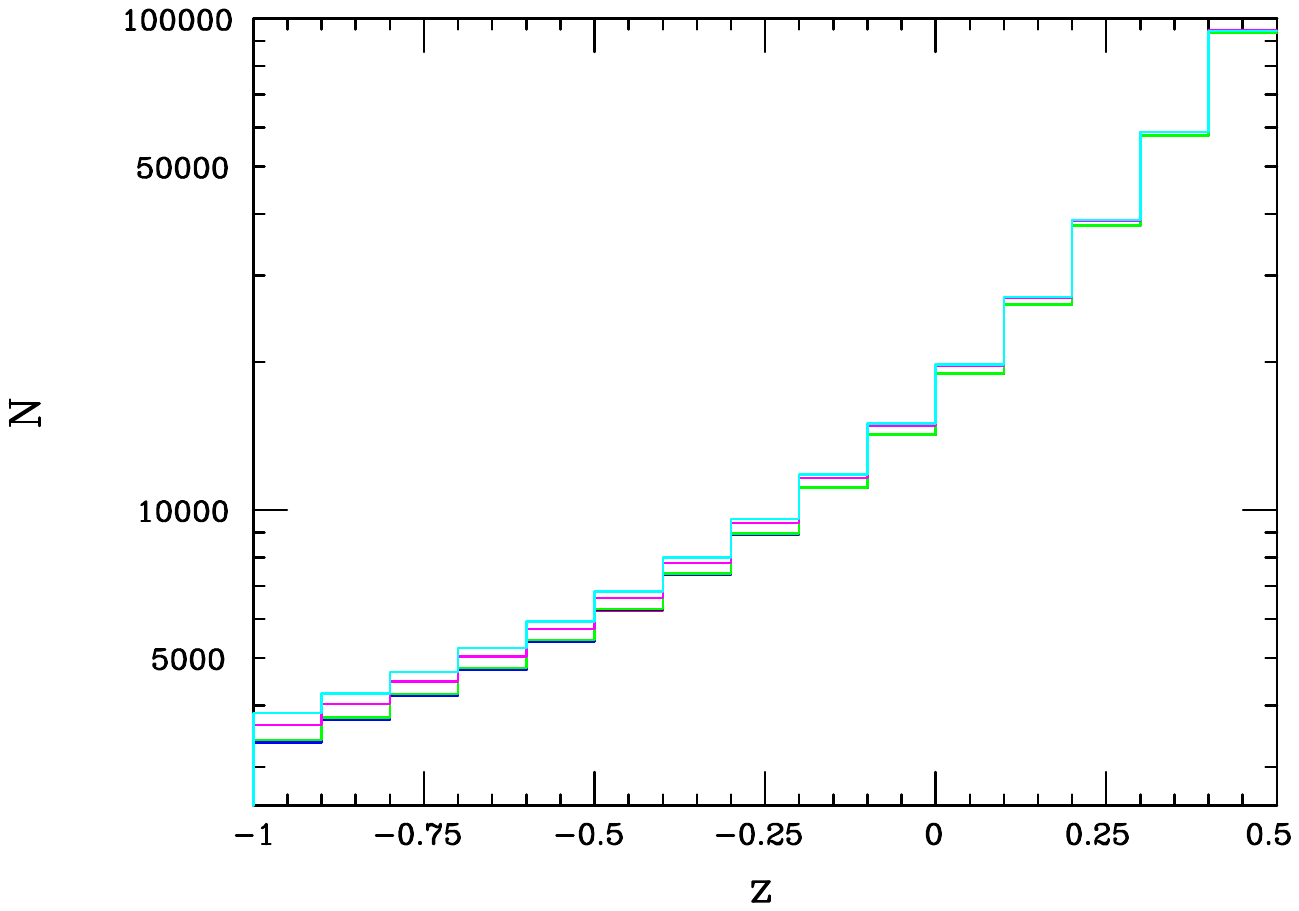}}
\vspace*{-1.3cm}
\caption{Same as in the Top panel in the previous Figure but now for Bhabha scattering and with $z\leq 0.5$,  away from the forward SM photon pole, so that some of the model dependence 
becomes easily visible.}
\label{fig10}
\end{figure}

Overall, it appears from this quick survey that $e^+e^-$ annihilation processes involving only SM fields will be able to probe the physics of the $M_U$ scale, at least for some range of parameters 
in our toy model which appears to overlap with, and may be somewhat larger than, a similarly successful region found in the case of DM production.

\section{Discussion and Conclusion}

The kinetic mixing scenario offers a well-motivated setup for sub-GeV scale thermal dark matter but can only be realized if new portal matter fields, either scalars or vector-like fermions, also exist 
which have charges under both the SM as well as the dark gauge group, $U(1)_D$, associated with the dark photon. However, once we posit the existence of such PM fields, it is easily imagined that 
they may lead to other additional mechanisms that can also generate SM-DM interactions, especially since one can argue that the $U(1)_D$ must be part of some larger non-abelian group, $G$, 
that breaks at a scale, $M_U$, that is likely not far away in energy, perhaps $\gsim$ a few TeV. One such possibility, here termed the DMom model, is to loop-generate form factor/dark moment-type 
three-point couplings for SM fermions to the DP in analogy with how a neutral Dirac neutrino or fermionic DM may interact with the SM photon. While the DMom and KM pictures both work well at 
rather low energies, if we want to probe physics at or above 
the $G$ breaking scale, which may lie somewhat beyond the reach of the HL-LHC to directly access in the form of new particle production, we are reduced to examining indirect signatures of such 
physics until much higher energy colliders, such as the FCC-hh, come on line in the somewhat distant future. If sub-GeV DM and a DP are discovered in nature and either of the KM or DMom setups 
are realized, measurements of the $e^+e^-\to V\gamma$ cross section by Belle II over the next few years should be able to differentiate between them. Not yet knowing the outcome of such 
measurements and positing that sub-GeV DM and DP states are {\it indeed} realized in nature, it behooves us to ask if indirect measurements made at intermediate energies below the $M_U$ scale, 
say between $M_Z$ and 1 TeV - the realm of future $e^+e^-$ colliders - can be used to tell us anything about the dynamics at/above such a scale. This is the question we've attempted to begin to address 
in the current paper and which we examined by employing simple toy model manifestations of these two distinct interaction frameworks of KM and DMom, here treated as being mutually exclusive.

Our first step was to identify the quantity or quantities in either setup that would probe this high scale physics. In the KM model, the energy evolution of the KM mixing parameter, $\epsilon$, is seen to 
arises from three distinct sources: those due to the RGE running of both of the $\alpha_D$ and $\alpha_{em}$ gauge couplings, which are controlled by either known SM or very low scale, dark 
sector physics that we expect will be fully explored by low energy measurements if 
this scenario is realized in nature, and the $M_U$ scale physics due to the PM fields. Unfortunately, as might be expected, we find in the context of $e^+e^-$ interactions that the highly suppressed 
nature of the KM-induced interaction between DM and SM fields necessary to achieve the observed relic density is maintained at higher energies until the $M_U$ scale is reached. This is true regardless 
of whether we are examining fermionic DM pair production or the effect of DP exchange between SM fields. So in this setup, we find that we are not able to probe any $M_U$ scale physics until the 
new states existing at such a mass can be produced on-shell at a future collider, \eg, FCC-hh. In the DMom scheme, on the other hand, the interactions are controlled by higher dimension operators 
so that, though suppressed at low energies to give the correct DM relic density, grow in strength with increasing $\sqrt s$ until roughly $\sim M_U$ is reached. Furthermore, only the leading term in the 
dark magnetic dipole-like coupling, $f_{20}$, enters into the low energy observables such as the relic density, direct detection searches, DP production in fixed target experiments or the 
$e^+e^-\to V\gamma$ process at Belle II. The next to leading order terms in the dark dipole moment coupling, $f_{21}$, as well as the corresponding leading term of the same dimension in the dark 
charge form factor, $f_1$, can only possibly make their presence known as collision energies increase into the few hundred GeV to TeV range as they both correspond to, effectively, higher dimensional 
operators, growing in strength with an addition power of $s$ relative to $f_{20}$. Given the knowledge of $f_{20}$ from the low energy measurements, it is only via the observation (and measurement ) 
of these two higher order terms that any information about $M_U$ scale physics be gleaned below this scale from the data. Thus we must not only be able to observe the necessary signals but also be 
able to determine that these higher dimensional operators are making a measurable contribution.

In this DMom setup, for the case radiative DM pair production, we find that the greatest difficulty is overcoming the very large but well-known SM background due to radiative neutrino production. While 
this background can be partially suppressed and the signal cross section can be somewhat enhanced by the same choice of beam polarizations, it is insufficient to allow for the observation of the DM signal 
for some regions of the toy model parameter space. However, there are other specific regions of this parameter space, particularly those where the PM and that of the non-hermitian gauge boson 
of $G$ within the toy model have similar masses, where the DM signal is visible and which allows the contributions of both $f_{21}$ and $f_1$ to be observable. In the case of unpolarized 
$e^+e^\to \mu^+\mu^-$, we seek the interference of the SM and dim-8 DP $s$-channel exchanges and we find that for similar (but potentially larger) parameter regions as in the case of radiative 
DM production the contribution of $f_1$ should be measurable although we find that the angular dependent beam polarization asymmetry is not very useful discriminator here. Lastly, for the case 
of unpolarized Bhabha scattering, the large event rate due to the photon pole in the forward direction more than compensates for the resulting reduced sensitivity to the new physics in the toy model and 
so we find greater sensitivity to $f_1$ than in the case of the dimuon final state. This was not completely unexpected as a similar result was obtained for the case of ADD gravity which also expresses 
itself at low energies as a dim-8 contact interaction. Thus, at least for the simple toy model example, we find that there do exist regions of the parameter space which allow us to indirectly probe 
$M_U$-scale physics through the use of high energy $e^+e^-$ colliders via the production of both DM as well as SM final states via DP exchange.

Finally, it goes without saying that if higher energy, $\sim 10$ TeV, lepton (either muon or $e^+e^-$) colliders were to eventually come on-line, much of the present analysis could easily be 
repeated with the appropriate scaling of the necessary integrated luminosity, except, of course, for the much larger machine-induced backgrounds which are to be expected and so would warrant 
a more detailed study, likely at the detector-level. However, once the $\sim 10$ TeV scale is indeed reached, it would be much more likely that the lepton-like PM fields participating in the loops 
could actually be directly pair-produced on-shell so that a different (and simpler) type of analysis would be needed to study their properties and would then yield more direct information about 
the physics at this scale. If the PM were to be even more massive, a detailed study of the potential $EE^*\to EeV$ final state would likely give access to PM masses even in excess of 
$\sim (0.90-0.95)\sqrt s$.

Hopefully signals of both dark matter and dark photons will soon be observed.

%------------------------------------ ACKNOWLEDGEMENTS ---------------------------------------%
\section*{Acknowledgements}
The author would like to particularly thank J.L. Hewett for valuable discussions and Brookhaven National Laboratory for its hospitality.  This work was supported by the Department of Energy, Contract 
DE-AC02-76SF00515.

%------------------------------------------- REFERENCES -------------------------------------------%

%-------------------------------------------------- END --------------------------------------------------%

\end{document}